\documentclass[12pt]{article}   

\makeatletter
\@addtoreset{equation}{section}
\makeatother

\def\be{\begin{equation}}
\def\ee{\end{equation}}
\def\ba{\begin{eqnarray}}
\def\ea{\end{eqnarray}}
\def\Nl{{\mathchoice
{\setbox0=\hbox{$\displaystyle\rm N$}\hbox{\hbox to0pt
{\kern0.4\wd0\vrule height0.9\ht0\hss}\box0}}
{\setbox0=\hbox{$\textstyle\rm N$}\hbox{\hbox to0pt
{\kern0.4\wd0\vrule height0.9\ht0\hss}\box0}}
{\setbox0=\hbox{$\scriptstyle\rm N$}\hbox{\hbox to0pt
{\kern0.4\wd0\vrule height0.9\ht0\hss}\box0}}
{\setbox0=\hbox{$\scriptscriptstyle\rm N$}\hbox{\hbox to0pt
{\kern0.4\wd0\vrule height0.9\ht0\hss}\box0}}}}
\def\Zl{{\mathchoice
{\setbox0=\hbox{$\displaystyle\rm Z$}\hbox{\hbox to0pt
{\kern0.4\wd0\vrule height0.9\ht0\hss}\box0}}
{\setbox0=\hbox{$\textstyle\rm Z$}\hbox{\hbox to0pt
{\kern0.4\wd0\vrule height0.9\ht0\hss}\box0}}
{\setbox0=\hbox{$\scriptstyle\rm Z$}\hbox{\hbox to0pt
{\kern0.4\wd0\vrule height0.9\ht0\hss}\box0}}
{\setbox0=\hbox{$\scriptscriptstyle\rm Z$}\hbox{\hbox to0pt
{\kern0.4\wd0\vrule height0.9\ht0\hss}\box0}}}}
\def\Ql{{\mathchoice
{\setbox0=\hbox{$\displaystyle\rm Q$}\hbox{\hbox to0pt
{\kern0.4\wd0\vrule height0.9\ht0\hss}\box0}}
{\setbox0=\hbox{$\textstyle\rm Q$}\hbox{\hbox to0pt
{\kern0.4\wd0\vrule height0.9\ht0\hss}\box0}}
{\setbox0=\hbox{$\scriptstyle\rm Q$}\hbox{\hbox to0pt
{\kern0.4\wd0\vrule height0.9\ht0\hss}\box0}}
{\setbox0=\hbox{$\scriptscriptstyle\rm Q$}\hbox{\hbox to0pt
{\kern0.4\wd0\vrule height0.9\ht0\hss}\box0}}}}
\def\Rl{{\mathchoice
{\setbox0=\hbox{$\displaystyle\rm R$}\hbox{\hbox to0pt
{\kern0.4\wd0\vrule height0.9\ht0\hss}\box0}}
{\setbox0=\hbox{$\textstyle\rm R$}\hbox{\hbox to0pt
{\kern0.4\wd0\vrule height0.9\ht0\hss}\box0}}
{\setbox0=\hbox{$\scriptstyle\rm R$}\hbox{\hbox to0pt
{\kern0.4\wd0\vrule height0.9\ht0\hss}\box0}}
{\setbox0=\hbox{$\scriptscriptstyle\rm R$}\hbox{\hbox to0pt
{\kern0.4\wd0\vrule height0.9\ht0\hss}\box0}}}}
\def\Cl{{\mathchoice
{\setbox0=\hbox{$\displaystyle\rm C$}\hbox{\hbox to0pt
{\kern0.4\wd0\vrule height0.9\ht0\hss}\box0}}
{\setbox0=\hbox{$\textstyle\rm C$}\hbox{\hbox to0pt
{\kern0.4\wd0\vrule height0.9\ht0\hss}\box0}}
{\setbox0=\hbox{$\scriptstyle\rm C$}\hbox{\hbox to0pt
{\kern0.4\wd0\vrule height0.9\ht0\hss}\box0}}
{\setbox0=\hbox{$\scriptscriptstyle\rm C$}\hbox{\hbox to0pt
{\kern0.4\wd0\vrule height0.9\ht0\hss}\box0}}}}
\def\Hl{{\mathchoice
{\setbox0=\hbox{$\displaystyle\rm H$}\hbox{\hbox to0pt
{\kern0.4\wd0\vrule height0.9\ht0\hss}\box0}}
{\setbox0=\hbox{$\textstyle\rm H$}\hbox{\hbox to0pt
{\kern0.4\wd0\vrule height0.9\ht0\hss}\box0}}
{\setbox0=\hbox{$\scriptstyle\rm H$}\hbox{\hbox to0pt
{\kern0.4\wd0\vrule height0.9\ht0\hss}\box0}}
{\setbox0=\hbox{$\scriptscriptstyle\rm H$}\hbox{\hbox to0pt
{\kern0.4\wd0\vrule height0.9\ht0\hss}\box0}}}}
\def\Ol{{\mathchoice
{\setbox0=\hbox{$\displaystyle\rm O$}\hbox{\hbox to0pt
{\kern0.4\wd0\vrule height0.9\ht0\hss}\box0}}
{\setbox0=\hbox{$\textstyle\rm O$}\hbox{\hbox to0pt
{\kern0.4\wd0\vrule height0.9\ht0\hss}\box0}}
{\setbox0=\hbox{$\scriptstyle\rm O$}\hbox{\hbox to0pt
{\kern0.4\wd0\vrule height0.9\ht0\hss}\box0}}
{\setbox0=\hbox{$\scriptscriptstyle\rm O$}\hbox{\hbox to0pt
{\kern0.4\wd0\vrule height0.9\ht0\hss}\box0}}}}

\def\k{\check}

\title{Coherent States for Canonical Quantum General Relativity and the 
Infinite Tensor Product Extension}
\author{H. Sahlmann\thanks{sahlmann@aei-potsdam.mpg.de},
T. Thiemann\thanks{thiemann@aei-potsdam.mpg.de},
O. Winkler\thanks{winkler@aei-potsdam.mpg.de} \\
       MPI f. Gravitationsphysik, Albert-Einstein-Institut, \\
           Am M\"uhlenberg 1, 14476 Golm near Potsdam, Germany}

\usepackage{amssymb}

\begin{document}

\maketitle

\begin{abstract}
We summarize a recently proposed 
concrete programme for investigating the (semi)classical limit
of canonical, Lorentzian, continuum quantum general relativity in four
spacetime dimensions. The analysis is based on a novel set of coherent 
states labelled by graphs. These fit neatly together with an Infinite Tensor
Product (ITP) extension of the currently used Hilbert space. The 
ITP construction enables us to give rigorous meaning to the infinite
volume (thermodynamic) limit of the theory which has been out of reach so 
far. 
\end{abstract}

\section{Introduction}
\label{s1}

Canonical Quantum General Relativity has matured over the last decade
into a serious candidate theory of quantum gravity which is manifestly
background independent. The most important developments include 
a rigorously defined mathematical framework, non-perturbative  
operator regularization methods, the prediction of a discrete or 
distributional (rather than smooth) quantum geometry at Planck scale, 
a microscopic explanation for the Bekenstein-Hawking black hole entropy 
and a fully diffeomorphism covariant state sum formulation. See \cite{1.2}
for pedagogical reviews, more or less covering these developments.

Recently, some effort \cite{1.1} has also been devoted to the extraction 
of semi-classical physics 
from the non-perturbative quantum theory. By this one means that
if the distances we probe are large compared to the Planck scale
$\ell_p=\sqrt{\hbar\kappa}$ ($\kappa$ is Newton's constant up to
a numerical factor in units where $c=1$) then the fluctuations of 
geometry can be neglected and we are close to a situation that one can
treat with the methods of quantum field theory on curved spacetimes,
which in turn also is valid only as long as the backreaction of geometry 
to quantum matter can be discarded.
In particular, if the gravitational field is in a semi-classical state
approximating the Minkowski metric then we should arrive at the quantum
field theory describing the standard model plus corrections.

What is a semi-classical gravitational state ? It is not entirely straight
forward to answer this question 
as the manifest background independence of the quantization
scheme forces us to adopt non-standard
representations of the canonical commutation relations, in particular,
the powerful Fock Hilbert space techniques are not available which would
immediately allow us to construct the standard coherent states used in the 
literature.
In what follows, the contents of a programme based on a new set of 
gravitational coherent states \cite{1.3} is summarized which 
provides a proposal for a systematic construction of the classical limit. \\
\\
The organization of the article is as follows :\\
\\
In section two we describe in detail and in a non-technical way 
the main ideas of our proposal which is based on a new class of 
kinematical coherent states, that is, minimal uncertainty states for {\it
both} connection and electric field operators labelled by a points in the 
classical, not necessarily constrained or reduced, phase space. Here 
we use the gauge theory canonically conjugate coordinates for this phase 
space described in \cite{1.2}. As such, they are in sharp contrast to the 
weave states already constructed in the literature which are 
are electric field operator 
eigenstates and thus completely outspread in the connection operator.
For every classical observable we construct 
an approximate (effective) quantum operator which has the correct 
expectation values and fluctuations with respect to our coherent states.
The term ``effective'' refers to the fact that these operators are, in 
general, different from those already constructed in the literature.
This poses no problem because to say that certain operator algebras 
correspond to certain classical Poisson algebras is an empty statement
without specifying with respect to which semi-classical states the 
correspondence is made. Our coherent states are simply not always adapted 
to the operators already constructed but we will point out how they can be 
modified in that respect.\\
We will 
see that three different scales emerge in the semi-classical analysis, 
the microscopic Planck scale, a mesoscopic scale defined by the edge length 
of an embedded graph (as measured by the three metric defined by a point in 
the classical phase space) and a macroscopic scale associated with the 
curvature of the four metric of that phase space point. When we ask 
that fluctuations be minimized we learn that the mesoscopic scale is 
a geometric mean of the other two scales. This result can be interpreted  
in the following way : Since the Planck length is proportional to 
a positive power of $\hbar$ and if we understand the classical limit roughly
as the limit $\hbar\to 0$ while the continuum limit corresponds
to $\epsilon\to 0$, then we may say that {\it the continuum and classical
limit merge into only one limit} ! More precisely, since $\hbar$ is a small
but finite number, also $\epsilon$ is a small but finite number and the 
actual continuum limit $\epsilon\to 0$ is unphysical, however, in an  
expansion in powers of $\epsilon$ the zeroth order term gives the correct
classical continuum expression.\\ 
An important step needed is the combination of model coherent states 
labelled by points in a model phase space (the cotangent bundle over 
a compact gauge group)
with the continuum phase space that we are actually interested in.
We discuss advantages and shortcomings of our proposal at the end of 
section \ref{s2} and we supplement it by appendices \ref{sa}, \ref{sb}
where we exhibit the derivation of an approximate version of the area 
operator and the appearance of optimal scales in fluctuation calculations. 

In section three we sketch the construction of model coherent
states in a hopefully pedagogical way, bearing, as much as possible
on an analogy with the harmonic oscillator coherent states of which
the present ones are a non-Abelean version. In particular, we try to provide some 
intuition based on geometric quantization and heat kernel methods.

In section four we introduce a technical tool, the Infinite Tensor 
Product construction which is well known in statistical
physics, in order to tackle the infinite volume limit of the theory
which has been out of reach so far. We outline just the most basic 
mathematical definitions and discuss how these mathematical termini translate
into physical concepts. This section is supplemented by appendix \ref{sc}
which employs the mathematical notions used in a simple system, the infinite
spin chain.

In the final section we summarize and list future applications of 
the concepts introduced, one of the most important being how to make contact 
with quantum field theory on curved spacetimes.

\section{A Concrete Programme for the Construction of (Semi)classical 
Canonical Quantum General Relativity}
\label{s2}

To say that a state is semi-classical makes sense only if one 
specifies a set of elementary observables $g_e$ ($e$ belongs to some 
label set) which are to behave almost
classically, not every possible operator on the Hilbert space
can have a good classical limit in the sense of approximating a given 
function on the classical phase space (see the third reference in \cite{1.1}
for an illustration in the context of canonical quantum general relativity).
The specification of a such a subset of the operator algebra is sometimes
referred to as {\it coarse-graining}.
One starts with a 
classical phase space $\cal M$ which in general depends also 
on the three -- manifold $\Sigma$ underlying the canonical formulation,
for example through boundary or fall-off conditions or through the 
differentiable structure (we refrain from displaying this dependence of
$\cal M$ on $\Sigma$ in what follows). 
A state $\psi_m$ labelled by a point $m$ in this classical 
phase space (say a canonical pair consisting of a connection $A_a^j$
and an electric field $E^a_j$) is said to be 
semi-classical for a set of elementary classical observables 
$g_e$ if the expectation 
values of the corresponding self-adjoint operators assume their 
classical 
values at the point $m$ and if their relative fluctuations are small,
that is, $<\psi_m,\hat{g}_e\psi_m>=g_e(m)$ and
\be \label{1.1}
(\Delta g_e)_{quant}(m)
:=\sqrt{<\psi_m,[\hat{g}_e]^2\psi_m>/[g_e(m)]^2-1}\ll 1
\ee
Clearly, as remarked by Ashtekar, if 
$g_e(m)=0$ then (\ref{1.1}) blows up but this is just because 
the value zero is smaller than the fundamental absolute fluctuation of the 
operator which is bounded away from zero. Nobody would say that the harmonic
oscillator coherent states break down at the origin of the phase space. If 
we replace $g_e$ by the quantity $g_e+c$ (which classically 
captures the same information as $g_e$) for some constant $c$ much
larger than the absolute fluctuation then (\ref{1.1}) is finite and small.
Alternatively, we can work with dimensionless quantities $g_e$ and consider
only absolute fluctuations. In what follows we assume that one has 
taken care of these subtleties in either way and that (\ref{1.1}) is 
well defined. Notice that the four -- metric determined by the initial data 
set $m$ introduces a first length scale into the analysis, namely its
typical curvature radius $L(m)$ as measured by the three metric determined
by $m$. As we will see in appendix \ref{sb} there are actually two such
radii (an electric and magnetic one) but we will symbolize them by a  
single label $L(m)$ for the descriptional purposes of this section.
If needed, we may want to restrict the phase space by allowing only those 
points $m$ for which $L(m)\ge L_0$ is bounded from below which could
induce a restriction on the (Gauss, diffeomorphism and Hamiltonian)
gauge choice involved in specifying $m$.

The states of the gravitational Hilbert space that we will consider are 
labelled by piecewise analytic graphs $\gamma$ with an at most countably 
infinite number of edges. These graphs are embedded into the three 
dimensional spatial manifold $\Sigma$ of given topology which underlies 
the canonical formulation classical general relativity.
For a given state it is only at the locus of the corresponding embedded 
graph where the excitations of the gravitational
field are probed. This raises immediately the question : What are these
degrees of freedom, called graph degrees of freedom in what follows, 
that are probed ? As far as the configuration degrees
of freedom are concerned, the answer is easy : they are the holonomies
of the gravitational connection along the edges of the graph. However,
for the momentum degrees of freedom this is less obvious. There is some 
freedom in choosing them and we will specify our choice below. Taken 
together, the observables associated with these degrees of freedom 
for  given graph $\gamma$
are good candidates for a set of elementary observables $g_e$ with 
respect to which a state $\psi_{\gamma,m}$, now also labelled by the 
graph, is semi-classical and $e$ will run through the set of edges of 
the graph. 

Now the following problem appears : The above comments suggest to
define semi-classical states over one single arbitrary but fixed graph 
$\gamma$. However, an arbitrary classical observable 
$O$ generically cannot be written as a function of the elementary graph 
degrees of freedom $g_e$ for the given graph $\gamma$ and therefore it
is not guaranteed that the corresponding continuum quantum operator 
will display semi-classical behaviour in the state $\psi_{\gamma,m}$.
A way out is suggested by the observation that any $O$
can be approximated by a function $O_\gamma=O_\gamma(\{g_e\})$ of the graph
degrees of freedom in the sense that 
\be \label{1.2}
(\Delta O)_{class}(m):=|\frac{O_\gamma(m)}{O(m)}-1|\ll 1
\ee
for all $m\in {\cal M}$. This can be seen as follows : Notice that given 
$\gamma,m$ automatically a second length scale $\epsilon$ is introduced,
namely the typical length (measured by the three metric determined by $m$)
of an edge of $\gamma$. Since $L$ is the scale on which the four metric 
varies, the classical approximation will be good as long as $\epsilon\ll L$.
If $L\ge L_0$ we just need to choose $\epsilon$ small enough to obtain
a small value of (\ref{1.2}) even uniformly in $m$. 

This observation motivates to use the corresponding operators 
$\hat{O}_\gamma$ as substitutes for the continuum operator $\hat{O}$.
There are two worrysome issues about this proposal : \\
I.)\\ 
The operators $\hat{O}_\gamma$ will, in general, not coincide with the 
cylindrical projections of the continuum operators already constructed 
in the literature, say the area operators \cite{1.4}. Thus, even if one 
manages to produce diffeomorphism covariant, cylindrically consistently 
defined families of operators 
$\hat{O}_\gamma$ corresponding to $O$ one will get a different continuum 
operator this way.\\
II.)\\
The substitute function $O_\gamma$ necessarily must
make explicit use of a coordinate system. Thus, unless one is very careful
in constructing it, it is conceivable that one breaks diffeomorphism
covariance and background independence. This would be  unsatisfactory 
because one of the strongest motivations and 
guidelines for this whole approach to quantum gravity is to keep background
independence at every stage of the construction. To see how this might 
happen, consider the so-called ``staircase problem'' which we discovered
in discussions with Ashtekar, Lewandowski and Pullin and
which we discuss more explicitly in appendix \ref{sa} : 
Suppose we would like to measure the area $O=A_S$ of a surface
$S$ in the state $\psi_{\gamma,m}$ where $\gamma$ is topologically a cubic 
lattice. If the surface is built from the elementary surfaces $S^e$ that come 
with the graph degrees of freedom as specified below then the cylindrical
projection of the 
operator constructed in \cite{1.4}, which is manifestly diffeomorphism
covariant, actually coincides with $\hat{O}_\gamma$ and its expectation
values agree with the classical values. However, if it lies transversally to 
them then its expectation value will be 
drastically off the classical value $A_S(m)$. The replacement $A_{S,\gamma}$ 
will have to take this error into account by using a coordinate system 
which enables us to tell how the elementary surfaces are embedded 
relative to $S$. (This coordinate system, however, does not depend on the 
point $m$). Now, while $g_e$ transforms covariantly under diffeomorphisms, 
there could be an extra coordinate dependence which is guaranteed to 
disappear only in the limit $\epsilon\to 0$ and $O_\gamma$ might not 
transform covariantly any longer.
Thus, the operator $\hat{O}_\gamma$ could be not only state dependent 
(it depends explicitly on $\gamma$ but not on $m$) but could also be 
coordinate dependent.
The former dependence can be removed if one can show that the 
$\hat{O}_\gamma$ are cylindrical projections of one and the same operator
$\hat{O}$ which can always be achieved by introducing suitable
projections, see below. However, it is possible that the extra coordinate 
dependence cannot be removed, at least not obviously, at any finite 
$\epsilon$. 

Of course, one could ignore this problem altogether 
and measure only operators that actually can be written in terms of the 
$\hat{g}_e$. In fact, as pointed out by Lewandowski, if one agrees that all 
physical information can be extracted, up to a certain accuracy set by a 
length scale, from electric and magnetic
fluxes through surfaces of the size of that scale or bigger then, up to a 
neglible
error, one can actually write these quantities purely in terms of the
$g_e$ due to the Gauss law and the Bianchi identity, see appendix \ref{sb}, 
in a manifestly coordinate independent way. While this is a working
proposal, we want to be more ambitious and measure any operator quantum
mechanically and not by first measuring elementary operators quantum 
mechanically and then assembling the values of these measurements by a 
classical formula into a value for the observable that we are interested 
in. Thus we have to deal with both problems I.), II.) in what follows.

We will show below that one can actually solve problem II.). Then one 
would still have problem I.), that is, for every classical observable $O$ 
one has two candidate quantum observables $\hat{O}',\hat{O}$, the first one 
of the kind already constructed in the literature using background 
independent regularization techniques
and the second one being constructed by the method outlined
below. One may object that given a classical function $O$ there must be a 
unique operator $\hat{O}$ such that its classical limit is given by $O$ and 
not several of them. However, in order to tell that a quantum theory given by a 
Hilbert space and a quantum commutator algebra has a given phase space
and Poisson algebra as its classical limit one needs one more input :
a selection of semi-classical states with respect to which the semi-classical
limit is to be performed. It is perfectly possible to have two systems of 
operators and two systems of semi-classical states such that the classical
limit of the first set of operators with respect to the first system of 
states agrees with the classical limit of the second set of operators with 
respect to the second system of states. There is no claim that this will 
continue to hold after exchanging the two systems of states. In fact,
our current situation in quantum general relativity is the following :
What we have are candidate operators $\hat{O}'$ already constructed in the 
literature and we have candidate semi-classical states $\psi$ constructed in 
\cite{1.3}. The staircase problem alluded to above reveals that these
states are not always semi-classical for the operators $\hat{O}'$. Thus
there exist two avenues to arrive at a satisfactory semi-classical
analysis :\\
Either A) one keeps the operators $\hat{O}'$ and modifies the states $\psi$
to arrive at better behaved semi-classical states $\psi'$ or B) one keeps the 
states $\psi$ and modifies the operators $\hat{O}'$ to arrive at better 
behaved operators $\hat{O}$. These are complimentary programmes aiming
at the same goal whose methods can be fruitfully combined so that one meets 
somewhere in the middle. 
The advantage of avenue A) is that the $\hat{O}'$
are explicitly known while the $\psi'$ are not, for 
avenue B) the $\psi$ are explicitly known while the $\hat{O}$ have to
be constructed graph by graph. \\
\\
In this paper we will mainly focus on avenue B) but we will devote this 
short paragraph to describe avenue A) in some detail in order to reveal
that both avenues can actually be combined.\\ 
Currently, avenue A) is being studied in the context of the   
{\it statistical geometry approach} \cite{1.5}. The idea is to construct 
semiclassical states without making reference to a particular, arbitrary
but fixed, graph $\gamma$ or any additional extra structure at all.
The state should be constructed just
from one datum, the phase space point $m$, in such a way that a sufficiently
coarse-grained and complete subset of the set of the operators $\hat{O}'$
already constructed in the literature behaves semi-classically.
Since these operators are manifestly diffeomorphism covariantly defined,
problem II.) never arises.
In order to remove effects similar to the staircase problem
which are associated with a direction dependence on a given graph, one 
obviously has to sum or average over a huge number of graphs. One could 
consider
either pure or mixed states but at this stage it is more natural to consider  
mixed states because then we do not need to choose the phases of 
the probability amplitudes involved but only their absolute values.
Choosing this option, instead of looking at one pure 
coherent state $\psi_{\gamma,m}$ one constructs a mixed state by averaging the 
one-dimensional projections $\hat{P}_{\gamma,m}$ onto $\psi_{\gamma,m}$ 
over a subset $\Gamma_m$ of the set of all allowed graphs (say piecewise 
analytic, compactly supported) which may depend on $m\in {\cal M}$ 
with a probability measure $d\mu_m(\gamma)$ on $\Gamma_m$. 
Here, one can be very general to begin with so that  $\psi_{\gamma,m}$ 
is not necessarily one of our coherent states, see \cite{1.5}.
Thus, one tries to construct a density matrix (a trace class operator 
of unit trace)
\be \label{1.5}
\hat{\rho}_m:=\int_{\Gamma_m} d\mu_m(\gamma) \hat{P}_{\gamma,m}
\ee 
and now expectation values are given by $<\hat{O}'>_m:=
\mbox{Tr}(\hat{\rho}_m\hat{O}')$ where $\hat{O}'$ is an operator on the 
currently used Hilbert space and not necessarily 
a function of the $\hat{g}_e$. Notice that, if as in \cite{1.5} the measure 
$\mu_m$ is absolutely continuous with respect to a Lebesgue measure, then
as an operator on the continuum 
Hilbert space, $\hat{\rho}_m$ is (almost) the zero operator (and has actually
zero trace) due to the 
diffeomorphism invariance of the inner product, however, the idea is 
to construct a new representation ${\cal H}_m$ of the operator algebra via 
the GNS construction from this state. This should be compared with
the strong equivalence class Hilbert spaces based on pure coherent states 
as cyclic vectors that we will discuss in section \ref{s4}. \\
Now, if the set $\Gamma_m$ is large enough then due to the averaging 
performed in 
(\ref{1.5}) the above pointed out direction dependence (staircase problem)
disappears. In fact, the explicit choice made in \cite{1.5} for 
flat initial data $m$ and compact $\Sigma$ is such that $\hat{\rho}_m$ is 
{\it Euclidean invariant} (using a weave state \cite{1.5a} or one of 
our coherent states for $\psi_{\gamma,m}$). 
The interested reader is referred to this beautiful statistical geometry 
construction
(based on the Dirichlet-Voronoi method \cite{1.6} for Euclidean spatial 
metrics) 
for more details. In general, there is certainly a lot of freedom in the 
choice 
of $\Gamma_m,\mu_m$ and it has to be understood how $<\hat{O}'>_m$ depends
on these choices. 

However, whatever choice is made, our coherent state 
construction fits neatly in with the idea to construct density matrices as
is obvious from the general formula (\ref{1.5}). For instance, using the 
peakedness properties derived in the third reference of \cite{1.3} it 
is easy to see that for flat initial data the coherent state is sharply 
peaked in the momentum representation {\it exactly at the value of the 
spin $j$ that has been previously derived in the weave construction} !
In other words, if we expand our coherent states in terms of spin-network
states, then practically all the coefficients except for the one with spin 
label $j_e\ell_p^2\approx \mbox{Ar}(S_e)$ are vanishing, where $e$ is an 
edge of $\gamma$, $S_e$ is a face 
of a polyhedronal decomposition of $\Sigma$ dual to $\gamma$ (see below)
which is intersected only by $e$ and Ar$(S_e)$ its area measured by the flat 
three-metric. This dual polyhedronal decomposition, an extra 
structure which is necessary to introduce in our coherent state construction
below and which comes with some freedom to choose from, {\it is actually 
naturally 
fixed by the Dirichlet Voronoi construction} ! This is a very surprising 
coincidence because the need for the polyhedronal decomposition in the
Dirichlet-Voronoi construction on the one hand and in the coherent 
state construction on the other hand are logically completely unrelated.
Roughly then, if a macroscopic surface $S$
is assembled from the $S_e$, the area expectation value can be practically 
taken with respect to a weave spin-network state at the above specified 
values of $j_e$ and gives the correct result already
reported earlier in the literature. If $S$ is not assmbled from the $S_e$
it still might bepossible, although no proof exists so far, that by using
some of the freedom just mentioned in our coherent state construction 
(specifically
in the the map $\Phi_{\k{\gamma},\Sigma,X}$ derived below) we can simply 
use our coherent states in this averaging method in order to arrive at a 
satisfactory semi-classical state. These issues are subject to future
research in collaboration with the authors of \cite{1.5}.\\
\\
Let us now come to avenue B) for which at this point one will work with a 
single, sufficiently large, gaph only rather than with an average.
Let us stress that restricting to a single graph is only due to mathematical
convenience and by no means crucial. 
In the future we might want to relax the condition to work with 
a single graph only and to use some kind of average over a {\it
countably infinite} number of graphs in the fashion of (\ref{1.5}).
The measure $\mu_m$ would in this case become a discrete (counting) measure 
which would not result in a precisely Euclidean invariant density operator
peaked on the Minkowski metric but this is maybe not bad because 
fundamentally there should not exist a 
precisely Euclidean invariant state in nature anyway if anything about  
the idea of a quantum spacetime foam at Planck scale is correct. On the other
hand, using a countable number of graphs in the average would have the 
advantage that we obtain a non-vanishing trace class operator of unit trace 
on the continuum Hilbert space and do not need to pass to a new GNS Hilbert 
space.

The easiest single graphs to use that come to one's mind are regular  
lattices (say of cubic topology). This has the obvious disadvantage 
that such graphs have preferred directions and do not look isotropic 
at any scale alrger than $\epsilon$. One idea to remove this direction 
dependence motivated by \cite{1.5} and suggested to us by Bombelli is to 
use for our $\gamma$ not a graph that is embedded as and/or is 
topologically a regular lattice but rather a (generic) {\it random graph} 
of the Dirichlet-Voronoi type so that there are no preferred directions 
on scales larger than or equal to an {\it isotropy scale} $I$ which in 
turn is much larger larger than $\epsilon$. As a result, 
in measuring macroscopic
observables (those that depend on a large number of elemnetary ones)
one does not detect the particulars of the graph any more. The isotropy 
scale $I$ gets absorbed in the particulars of the classical error
(\ref{1.2}) as we will see in appendix \ref{sb} and as a net effect 
influences the size of the curvature scale $L$. Having said this, 
we will not explicitly display the dependence of $L$ on $I$ any more in
this paper.

Our first task is to show that one can indeed solve problem II.), that is,
the $\hat{O}_\gamma$ have to be constructed in diffemorphism covariant 
fashion. To do this, recall that in the canonical approach to quantum 
gravity one assumes that
the four-dimensional manifold $M$ has topology $\Rl\times \Sigma$ where
$\Sigma$ is an analytic three-dimensional manifold of given topology. The 
graphs are unions of 
edges which themselves are one-dimensional analytic submanifolds of 
$\Sigma$ intersecting at most in their endpoints. Likewise, classical
functions $O_S$ on $\cal M$ typically depend on analytic submanifolds
$S$ of $\Sigma$ like curves, surfaces and regions. Also, as we will see 
below, in order to define the graph degrees of freedom we must actually
also specify a polyhedronal decomposition $P_\gamma$ dual to $\gamma$
and a set of paths $\Pi_\gamma$ inside the corresponding faces.
One can describe these structures
mathematically by introducing a coordinate system $X$ (to which 
we will refer as an {\it embedding} in what follows) and {\it model} 
graphs $\k{\gamma}$, faces $P_{\k{\gamma}}$, sets of paths $\Pi_{\k{\gamma}}$ 
and surfaces $\k{S}$ in the model space $\k{\Sigma}$
of $\Sigma$ (that is, $\Rl^3$) such that 
$\gamma=X(\k{\gamma}),P_\gamma=X(P_{\k{\gamma}}),
\Pi_\gamma=X(\Pi_{\k{\gamma}}),\;S=X(\k{S})$. In what follows, model objects
like these will carry a check sign as compared to their embedded 
counterparts. Given $\Sigma$, fix once and for all a coordinate system $X_0$.
Given a pair $\gamma,S$, choose once and for all a representant 
$(\gamma_0,S_0)$ in their orbit 
$[(\gamma,\Sigma)]:=\{(\varphi(\gamma),\varphi(S));\;
\varphi\in Diff(\Sigma)\}$ under diffeomorphisms and a diffeomorphism
$\varphi_{S,\gamma}$ such that 
$\varphi_{S,\gamma}(\gamma_0)=\gamma,\varphi_{S,\gamma}(S_0)=S$. Furthermore,
we require that $\gamma_0$ is the same graph for pairs $(\gamma,S)$,
$(\gamma',S')$ such that $\gamma,\gamma'$ are diffeomorphic. Notice that
$\gamma_0,S_0,\varphi_{S,\gamma}$ are far from unique and in order 
to choose them we make use of the axiom of choice. 

Define the model graph and surface $\k{\gamma},\k{S}$ 
respectively by $\gamma_0=X_0(\k{\gamma}),S_0=X_0(\k{S})$ and choose 
once and for all model faces and paths $P_{\k{\gamma}},\Pi_{\k{\gamma}}$
for $\k{\gamma}$. Pick a prescription to 
construct a model function 
$O_{\k{S},\k{\gamma},P_{\k{\gamma}},\Pi_{\k{\gamma}}}$, 
which depends only on the model graph degrees
of freedom $g_{\k{e}}$. The model graph degrees of freedom are labelled 
explicitly by a model edge $\k{e}$ (but no dual face and correspeonding 
path system $S^{\k{e}},\Pi_{\k{e}}$ respectively since we fixed them
once and for all) and take values in a
model phase space ${\cal M}_{\k{\gamma}}$ as we will see below.
By definition, the numerical coefficients that the model function
involves {\it do not explicitly depend either on 
$X_0$ or on $m$} ! Rather they depend only on 
$\k{\gamma},P_{\k{\gamma}},\Pi_{\k{\gamma}},\k{S}$ and 
possibly additional extra structures in $\k{\Sigma}$.
Now, given $X_0,\Sigma$ there is a natural map $\Phi_{\k{\gamma},\Sigma,X_0}$ 
from $\cal M$ into ${\cal M}_{\k{\gamma}}$ defined by
$g_{\k{e}}:=g_{e_0}(m)=:\Phi_{\k{\gamma},\Sigma,X_0}(m)$ where 
$g_{e_0}$ depends explicitly on the embedded structures
$e_0=X_0(\k{e}),S^{e_0}=X_0(S^{\k{e}}),\Pi_{e_0}=X_0(\Pi_{\k{e}})$ 
respectively. The point is now that if $O_S$ was diffeomorphism covariantly 
and background independently defined
then there is a natural way to construct this model function
in such a way that 
$O_{\k{S},\k{\gamma},P_{\k{\gamma}},\Pi_{\k{\gamma}}}\circ
\Phi_{\k{\gamma},\Sigma,X_0}$
is close to $O_{S_0}$ pointwise in $\cal M$
as we will explicitly demonstrate in appendices \ref{sa}, \ref{sb}. 

We can then finish
the construction of $O_{S,\gamma}$ through the definition
\be \label{1.1a}
O_{S,\gamma}(m)
:=O_{\k{S},\k{\gamma},P_{\k{\gamma}},\Pi_{\k{\gamma}}}
(\{\varphi_{S,\gamma}\cdot g_{e_0}(m)\}_{e_0\in E(\gamma_0)})
\ee
where $(\varphi\cdot g_e)(m):=g_e(\varphi^{-1}\cdot m)$ is the  
natural action of $Diff(\Sigma)$ on the $g_e$ induced by that on $\cal M$
and $E(\gamma)$ denotes the set of edges of $\gamma$.
Now it is easy to see that due to the diffeomorphism covariance of the
functions $g_e$ we actually have 
$\varphi_{S,\gamma}\cdot\Phi_{\k{\gamma},\Sigma,X_0}
=\Phi_{\k{\gamma},\Sigma,X_{\varphi_{S,\gamma}}}$
where $X_\varphi:=\varphi\circ X$. We can therefore write (\ref{1.1a})
more compactly as 
\be \label{1.1a1}
O_{S,\gamma}
:=O_{\k{S},\k{\gamma},P_{\k{\gamma}},\Pi_{\k{\gamma}}}
\circ\Phi_{\k{\gamma},\Sigma,X_{\varphi_{S,\gamma}}}
\ee
which shows we have automatically
a natural action of the diffeomorphism group on $O_{S,\gamma}$ given by
\be \label{1.1b}
(\varphi\cdot O_{S,\gamma})(m)
=O_{S,\gamma}(\varphi^{-1}(m))
\ee
as expected from a function on the classical phase space.
However, in general 
$\varphi\cdot O_{S,\gamma}\not=O_{\varphi(S),\varphi(\gamma)}$ in contrast
to $\varphi\cdot O_S=O_{\varphi(S)}$ for the exact classical function.
Since $\lim_{\epsilon\to 0} O_{S,\gamma}(m)=O_S(m)$ pointwise on 
$\cal M$ by construction, this behaviour under diffeomorphisms is 
expected only in the continuum limit although it will be close to it
for sufficiently small $\epsilon$. 
This violation of a continuum property of the classical function 
is the price to pay for making the function graph dependent. There 
are two complementary points of view, reached in  
a discussion with Ashtekar, Lewandowski and Pullin : Either one regards the 
approximate operators as effective ones, good only for a semi-classical
regime. Or one takes the opposite point of 
view, that since we want to quantize general relativity 
non-perturbatively and background-independently, we are forced to adopt 
a non-standard Hilbert space whose elements are labelled by graphs.
Therefore, the $O_{S,\gamma}$ are the {\it more fundamental building
blocks from which everything else must be derived} and as long as
a continuum property is regained in the limit of infinitely fine graphs,
this is acceptable. This issue will be explored in more depth in 
future publications.

As we will show below, all that is important about the map
$\Phi_{\k{\gamma},\Sigma,X_{\varphi_{S,\gamma}}}$ for quantization 
purposes is that $X_{\varphi_{S,\gamma}}(\k{\gamma})=\gamma$ so that
the correct quantization of the functions (\ref{1.1a1}) is given by
\be \label{1.1c}
\hat{O}_{S,\gamma}
:=O_{\k{S},\k{\gamma},P_{\k{\gamma}},\Pi_{\k{\gamma}}}
(\{\hat{g}_e\}_{e\in E(\gamma)})
\ee
Namely, we will show below that this operator, interpreted as an operator on 
the closed subspace ${\cal H}^\gamma$ of the full continuum Hilbert space
$\cal H$ spanned by spin-network functions over $\gamma$ \cite{1.7}
with dependence on every edge by non-trivial irreducible representations
of $SU(2)$, can be written in terms of holonomy operators $\hat{h}_e$
and momentum operators $\hat{P}^e_j$ acting by multiplication and
differentiation with respect to $h_e$ respectively where $e\in E(\gamma)$.
Then, interpreting $\hat{g}_e$ as the pair $(\hat{h}_e,\hat{P}^e_j)$ 
there is a natural quantum action of the diffeomorphism group
on these operators defined by
\be \label{1.d}
\hat{U}(\varphi)\hat{g}_e\hat{U}(\varphi)^{-1}=\hat{g}_{\varphi(e)}
\mbox{ and }
\hat{U}(\varphi)\hat{O}_{S,\gamma}\hat{U}(\varphi)^{-1}=
O_{\k{S},\k{\gamma},P_{\k{\gamma}},\Pi_{\k{\gamma}}}(\hat{g}_{\varphi(e)})
\ee
so that it is completely diffeomorphism covariantly defined.
Moreover, it leaves the subspace 
${\cal H}_\gamma
:=\overline{\oplus_{\gamma'\subset\gamma} {\cal H}^{\gamma'}}$
invariant (the sum running over subgraphs of $\gamma$). The corresponding
family of operators 
$\hat{O}_S:=\{\hat{O}_{S,\gamma}\}_{\gamma\in\Gamma}$ may not yet be 
consistently defined in the sense that 
$(\hat{O}_{S,\gamma})_{|{\cal H}_{\gamma'}}
=(\hat{O}_{S,\gamma'})_{|{\cal H}_{\gamma'}}$ for any 
$\gamma'\subset\gamma$. If that is not already  
the case we replace $\hat{O}_{S,\gamma}$ by 
$\sum_{\gamma'\subset\gamma}\hat{O}_{S,\gamma'}\hat{P}_{\gamma'}$ (or 
$\sum_{\gamma'\subset\gamma}\hat{P}_{\gamma'}\hat{O}_{S,\gamma'}
\hat{P}_{\gamma'}$ to make it explicitly self-adjoint)
where $\hat{P}_{\gamma'}$ is an orthogonal projection onto
${\cal H}_{\gamma'}$. From now on we assume that $\hat{O}_S$ has been 
consistently defined like this and thus we have arrived at 
a new definition of continuum operators $\hat{O}$ which are diffeomorphism 
covariantly defined without going through a regularization procedure.
This finishes our task associated with problem II.)\\
\\
In what follows, we will only work with this new kind of operators and 
drop the label $S$ for $O_S$ again, $O$ wil be a general classical
function on $\cal M$. Then the next question is : Given $\Sigma,m$, 
for which graph should one construct a semi-classical state 
in order to arrive at the desired semi-classical 
interpretation, in particular, what is the size of 
$\epsilon$ ? This is connected with the question of the continuum limit 
$\epsilon\to 0$. Certainly, $\epsilon$ should be small enough so that
the classical error (\ref{1.2}) is small. But there are also quantum 
errors : First of all, unless the corresponding quantum operator 
$\hat{O}_\gamma:=O_\gamma(\hat{g}_e)$ is normal ordered 
we also have a normal ordering error 
\be \label{1.3}
(\Delta O_\gamma)_{normord}(m):=
|\frac{<\psi_{\gamma,m},\hat{O}_\gamma\psi_{\gamma,m}>}{O_\gamma(m)}-1|
\ee
which in applications is usually of the same order as the 
the quantum fluctuation of $\hat{O}_\gamma$ given by 
\be \label{1.4}
(\Delta O_\gamma)_{quant}(m):=
<\psi_{\gamma,m},[\frac{\hat{O}_\gamma}{O_\gamma(m)}-1]^2
\psi_{\gamma,m}>^{1/2}
\ee
These quantum errors are not yet the quantum errors 
of our consistently defined continuum operator $\hat{O}$.
However, since when 
computing expectation values or fluctuations of $\hat{O}$ on cylindrical
functions we actually project $\hat{O}$ to one of the $\hat{O}_{\gamma}$
we can introduce the following notion of fluctuation of a real valued
observable $O$ for which we assume $\hat{O}_{\gamma}$ to be 
self-adjoint :
\be \label{1.7}
(\Delta O)_{total}(m):=
<\psi_{\gamma,m},[\frac{\hat{O}}{O(m)}-1]^2\psi_{\gamma,m}>^{1/2}
=<\psi_{\gamma,m},[\frac{\hat{O}_\gamma}{O(m)}-1]^2\psi_{\gamma,m}>^{1/2}
\ee
which measures the fluctuation of the substitute operator  
$\hat{O}_\gamma$
compared to the exact classical value $O(m)$. It is important to realize 
that due to cylindrical consistency we could replace $\hat{O}_{\gamma}$
in (\ref{1.7}) by $\hat{O}$. This quantity can be written
in terms of the classical, normal ordering and quantum error as 
\ba \label{1.8}
(\Delta O)_{total}(m)^2 &=&
[\frac{<\hat{O}_{\gamma}>}{O_\gamma(m)}]^2
[\frac{O_\gamma(m)}{O(m)}]^2 (\Delta \hat{O}_{\gamma})_{quant}(m)^2
\nonumber\\
&& +|\frac{O_\gamma(m)}{O(m)}(\frac{<\hat{O}_{\gamma}>}{O_\gamma(m)}-1)
+(\frac{O_\gamma(m)}{O(m)}-1)|^2
\ea
Equation (\ref{1.8}) allows us to derive the
following important inequality
\ba \label{1.9}
(\Delta O)_{total}(m) &\le&
[1+(\Delta O_{\gamma})_{normord}(m)]\;[1+(\Delta O)_{class}(m)]
(\Delta O_{\gamma})_{quant}(m)
\nonumber\\
&& +[1+(\Delta O)_{class}(m)](\Delta O_{\gamma})_{normord}(m)
+(\Delta O)_{class}(m)
\ea
and thus will be small if both $(\Delta O_\gamma)_{quant}(m)$ and 
$\epsilon/L$ are small. If practically possible and physically motivated 
one should of course normal order the operator $\hat{O}_\gamma$.
These fluctuations will be the smaller the more elementary observables, say 
$N$, $O_\gamma$ involves {\it additively} (see first reference of 
\cite{1.1}), that is, the more macroscopic $O$ is. This is because 
of the law of large numbers for fluctuation errors which decreases
as $1/\sqrt{N}$. The least macroscopic observables that we can measure are 
the elmentary ones, $O=g_e$ for which we require still good semi-classical
behaviour and which we will call the {\it mesoscopic} ones. We obviously
will not be allowed to choose $\epsilon$ as small
as $\ell_p$ or lower because then $(\Delta O_\gamma)_{quant}(m)$ 
will be large since we would probe the geometry at the {\it microscopic}
Planck scale.

These considerations suggest the relation $\ell_p\ll\epsilon\ll L(m)$
or $\ell_p\ll\epsilon\ll L_0$ if we want $\epsilon$ to be independent
of $m$. We think of $\epsilon$ as a parameter for graphs which we can 
adjust in such a way as to minimize fluctuations. We will see later in 
appendix \ref{sb} that minimization fixes $\epsilon$ to be of the order of  
$\epsilon=\ell_p^r L(m)^s$ or $\epsilon=\ell_p^r L_0^s$ 
where $r,s$ are positive rational numbers adding up to one.
Moreover, the right hand side of (\ref{1.9}) turns out to be 
(not surprisingly) of the 
form $(\ell_p/L(m))^\alpha\le (\ell_p/L_0)^\alpha$ for some positive 
rational number $\alpha$.
{\it We see that the continuum limit $\epsilon\to 0$ and the 
classical limit $\hbar\to 0$ are no longer separate limits but 
happen simultaneously}. This at least restricts our freedom
to choose a graph, given $\Sigma,m$ in order to enforce maximum 
semi-classical behaviour. 

We are then left with the specification of the degrees of freedom 
$g_{\k{e}}$ and the maps $\Phi_{\k{\gamma}, \Sigma, X}$, or equivalently,
the $g_e$. Once we have singled them out, we will choose the states 
$\psi_{\gamma,m}$
to be {\it coherent states} for the $\hat{g}_{e}$, that is, an overcomplete 
set of minimal uncertainty states for the $g_e$, see section
\ref{s3}.
Given an embedded graph $\gamma$ and point $m\in{\cal M}$ we simply
take the holonomies $h_e(A)$ along its edges as the configuration degrees
of freedom where $A$ is the connection datum specified by $m=(A,E)$. 
As already announced, in
order to specify momentum degrees of freedom we have to blow up the label 
set $\gamma$ somewhat : We consider a polyhedronal decomposition $P_\gamma$
of the 
three -- manifold $\Sigma$ such that the graph $\gamma$ is dual to it. Thus, 
for each edge $e$ of $\gamma$ there is a unique, open, face $S^e$ of the 
polyhedronal decomposition which it intersects transversally in an 
interior point $p_e$ of both $e$ and $S^e$ and whose orientation agrees 
with that of $e$. 
Next, for each $x\in S^e$ we choose a non-self-intersecting path $\pi_e(x)$ 
within $S^e$ which starts at $p_e$ and ends at $x$ without winding around 
any point in $S^e$. The collection of these paths will be denoted by 
$\Pi_\gamma$. Finally, let $a$ be a fixed parameter of the dimension
of a length. Then we define the dimensionless quantity
\be \label{1.10}
P^e_j(A,E):=-\frac{1}{2 a^2}\mbox{tr}(\tau_j\mbox{Ad}_{h_e(p_e)}
[\int_{S^e} \mbox{Ad}_{h_{\pi_e(x)}}(\ast E(x))])
\ee
where $h_e(p_e)$ is the holonomy along the segment of $e$ starting at 
the starting point of $e$ and ending at $p_e$, $\ast$ denotes the metric 
independent
Hodge dual and $E=E_j\tau_j$ with respect to generators $\tau_j$
of $su(2)$ with the properties $\mbox{tr}(\tau_j\tau_k)=-2\delta_{jk},
[\tau_j,\tau_k]=2\epsilon_{jkl}\tau_l$. Notice that (\ref{1.10}) is 
gauge covariant, that is, under local gauge transformations it transforms in 
the adjoint representation of $SU(2)$ at the starting point of $e$. It is 
also background independently and diffeomorphism covariantly defined.
Moreover, $P^e_j(m)$ depends linearly on the electric field datum $E$
specified by $m$. The interpretation of and the motivation for the quantitity 
$a$ with the dimension of a length is not clear 
at this point but it will become obvious later on when we compute 
fluctuations where it will be fixed at the order of $L(m)$ or $L_0$.
Also, tuning $a$ will tune the Gaussian width of the coherent states that 
we construct.

Together
with the holonomies $h_e(m)$ of the connection along edges we obtain the 
elementary graph degrees of freedom 
\be \label{1.10a}
g_{e,\Sigma,P_\gamma,\Pi_\gamma}(m):=(h_e(m),P^e_j(m)) 
\ee
which, abusing the notation, 
are specific functions on $\cal M$ that depend 
on $\Sigma,\gamma,P_\gamma,\Pi_\gamma$. It is 
therefore very surprising
at first that they satisfy a Poisson algebra (see the first reference in 
\cite{1.3}), derived from the canonical
Poisson brackets $\{.,.\}$ between the $A,E$ on $\cal M$, which is 
{\it completely independent of $\Sigma,P_\gamma,\Pi_\gamma$}, specifically
\ba \label{1.11}
\{h_e,h_{e'}\}(m)&=&0 
\nonumber\\
\{P^e_j,h_{e'}\}(m)&=&\frac{\kappa}{a^2}
\delta^e_{e'}\frac{\tau_j}{2}h_e(m)
\nonumber\\
\{P^e_j,P^{e'}_k\}(m)&=&-\delta^{e e'}\frac{\kappa}{a^2}
\epsilon_{jkl} P^e_l(m)
\ea 
With a little experience in background independent quantum field 
theory, however, this is not so surprising, it just mirrors the fact
that no metric, distances, angles etc. can appear in (\ref{1.11}).

Let us now forget about the structures $\Sigma,
P_\gamma,\Pi_\gamma$ altogether and consider a model graph
$\k{\gamma}$ which can be embedded as $\gamma$ into $\Sigma$.
Let us also forget the continuum phase space $({\cal M},\{.,.\})$
and rather consider a model phase space 
$({\cal M}_{\k{\gamma}},\{.,.\}_{\k{\gamma}})$ with coordinates 
$h_{\k{e}}\in SU(2),\;P^{\k{e}}_j\in \Rl^3$ which specify
$g_{\k{e}}=(h_{\k{e}},P^{\k{e}}_j)$ and the point 
$m_{\k{\gamma}}:=\{g_{\k{e}}\}_{\k{e}\in E(\k{\gamma})}$ where 
$E(\k{\gamma})$ is the set of directed edges of $\k{\gamma}$. These 
coordinates enjoy the basic brackets inherited from (\ref{1.11})
\ba \label{1.12}
\{h_{\k{e}},h_{\k{e}'}\}_{\k{\gamma}}&=&0 
\nonumber\\
\{P^{\k{e}}_j,h_{\k{e}'}\}_{\k{\gamma}}&=&\frac{\kappa}{a^2}
\delta^{\k{e}}_{\k{e}'}\frac{\tau_j}{2}h_{\k{e}}
\nonumber\\
\{P^{\k{e}}_j,P^{\k{e}'}_k\}_{\k{\gamma}}
&=&-\delta^{\k{e}\k{e}'}\frac{\kappa}{a^2}\epsilon_{jkl} P^{\k{e}}_l
\ea 
which defines the natural symplectic structure of the $|E(\k{\gamma})|$
fold copy of the cotangent bundle $T^\ast SU(2)$, displaying
${\cal M}_{\k{\gamma}}$ as the cotangent bundle over the 
$|E(\k{\gamma})|$ fold copy of $SU(2)$.

We now want to construct coherent states for ${\cal M}_{\k{\gamma}}$.
These will be states on the model Hilbert space (which is of course
isomorphic with ${\cal H}_\gamma$)
\be \label{1.13}
{\cal H}_{\k{\gamma}}:=\otimes_{\k{e}\in E(\k{\gamma})} L_2(SU(2),d\mu_H)
\ee
where $\mu_H$ is the Haar measure on $SU(2)$, that is, they will functionally
depend on the point $h_{\k{\gamma}}:=\{h_{\k{e}}\}_{\k{e}}$ of the 
configuration subspace of ${\cal M}_{\k{\gamma}}$. 
On the other hand, we want them to be 
peaked at a point $m_{\k{\gamma}}$ of the phase space $M_{\k{\gamma}}$
and therefore they will carry a label $m_{\k{\gamma}}$. The tensor product
structure of the Hilbert space makes it obvious that these states will 
be of the form of a direct product
\be \label{1.14}
\psi_{\k{\gamma},m_{\k{\gamma}}}(h_{\k{\gamma}})=\prod_{\k{e}} 
\psi^t_{g_{\k{e}}}(h_{\k{e}})
\ee
where $\psi^t_g(h)$ is a coherent state on the Hilbert space 
$L_2(SU(2),d\mu_H)$ with label $g\in T^\ast SU(2)$ functionally depending
on $h\in SU(2)$. These states depend explicitly on the dimensionless
{\it classicality parameter}
\be \label{1.15}
t=\frac{\ell_p^2}{a^2}
\ee
which naturally arises in quantizing the above Poisson bracket, e.g.
$[\hat{P}^{\k{e}}_j,\hat{P}^{\k{e}'}_k]=-it\delta^{\k{e}\k{e}'}
\epsilon_{jkl} \hat{P}^{\k{e}}_l$. It is important to see that the Hilbert 
space
(\ref{1.13}) carries a faithful representation of these commutation relations and 
the adjointness conditions 
$(\hat{P}^{\k{e}}_j)^\dagger=\hat{P}^{\k{e}}_j,\;
(\hat{h}_{\k{e}}^{AB})^\dagger=(\hat{h}_{\k{e}}^{BA})^{-1}$
following from the classical reality (unitarity) respectively if
$\hat{h}_{\k{e}}$ acts by multiplication and 
$\hat{P}^{\k{e}}_j$ is $it/2$ times the right invariant vector field 
on the copy of $SU(2)$ corresponding to $h_{\k{e}}$ generating left
translations in the direction of $\tau_j$ in $su(2)$.

More explicitly these states are labelled
by points in $SL(2,\Cl)$ which is possible since one can 
identify $g_{\k{e}}$ with
the following elemement of $SL(2,\Cl)$ (we abuse the notation in using the 
same symbol) 
\be \label{1.16}
g_{\k{e}}:=\exp(-i\tau_j P^{\k{e}}_j/2)h_{\k{e}}
\ee
which is actually a diffeomorphism from $T^\ast SU(2)$ to $SL(2,C)$ 
(the inverse map being given by polar decomposition). Since the former
is a sympletic manifold and the latter a complex manifold it is not
surprising that $SL(2,C)$ is actually a K\"ahler manifold (the complex
structure is compatible with the symplectic structure) which suggests to use
well known methods from geometric quantization (in particular, heat
kernel methods) for the construction of the coherent states. These 
methods will be described in some detail in section \ref{s3}, here it is 
sufficient to contemplate that the situation is actully quite analogous
to the harmonic oscillator coherent states which are labelled by a point
$z$ in the complex plane, whose real and imaginary part respectively
can be interpreted as configuration $q$ and momentum $p$ degree of freedom 
respectively, and which depend functionally on a point $x\in\Rl$ in the 
configuration space of the phase $T^\ast \Rl$. The complexification of the 
real line is the complex plane and the complexification of the group 
$SU(2)$ is given by $SL(2,\Cl)$. Thus, (\ref{1.16}) is nothing else than 
a non-Abelean and exponentiated version of $z$.

This finishes the construction of a coherent state on the model Hilbert 
space ${\cal H}_{\k{\gamma}}$ labelled by a point in a model
phase space. We now must make contact with the Hilbert space $\cal H$ and
the phase space $\cal M$ appropriate to describe a semi-classical situation 
for a given topology $\Sigma$. The formulae (\ref{1.10}),
(\ref{1.11}) provide the necessary clue : Given a model graph 
$\k{\gamma}$, a three manifold $\Sigma$,
an embedding $X$ of $\k{\gamma}$ into $\Sigma$ as $\gamma=X(\k{\gamma})$,
a polyhedronal decomposition $P_{\k{\gamma}}$ of $\Sigma$ dual to 
$\k{\gamma}$ 
and a choice $\Pi_{\k{\gamma}}$ of paths within its faces we can construct 
the concrete maps (\ref{1.10a}) which provide us with a map (we display
only its dependence on $\k{\gamma}$ but the dependence on the other 
structures should be kept in mind)
\be \label{1.17}
\Phi_{\k{\gamma},\Sigma,X}:\;{\cal M}\mapsto {\cal M}_{\k{\gamma}};\;
m\mapsto \{g_{\k{e}}:=g_{X(\k{e}),\Sigma,X}(m)\}_{e\in E(\k{\gamma})}
\ee
which, according to (\ref{1.11}), (\ref{1.12}), is actually a
symplectomorphism (more precisely, it leaves the Poisson brackets 
invariant but it is not invertible and not even necessarily surjective
if $\Sigma$ is not compact since then $m$ needs to obey certain fall-off
conditions). The map (\ref{1.17}) is the final ingredient to obtain
a semi-classical state appropriate for a given $\Sigma$ and phase space 
point $m$
\be \label{1.18}
\psi_{\gamma,m}(h_\gamma):=
\psi_{\k{\gamma},\Phi_{\k{\gamma},\Sigma,X}(m)}(h_\gamma)
\ee
where we have identified ${\cal H}_\gamma$ with ${\cal H}_{\k{\gamma}}$,
in particular, $h_e$ with $h_{\k{e}}$ whenever $X(\k{e})=e$.
Since the model Hilbert space is the same for any $\Sigma$, this opens the
possibility that one might be able to  
{\it to describe topology change within canonical quantum general 
relativity} : If we can define consistently a new representation of the 
canonical commutation relations based on model graphs rather than 
embedded ones, then the transition amplitude \\
$<\psi_{\k{\gamma},\Phi_{\k{\gamma},\Sigma,X}(m)},
\psi_{\k{\gamma},\Phi_{\k{\gamma},\Sigma',X'}(m')}>$
is well-defined since the scalar product is computed on the common model 
Hilbert space and not on the embedded one and one and the same model graph 
can be embedded differently
into different $\Sigma$. We elaborate more on this idea in future 
publications \cite{1.13} where methods from algebraic graph theory 
will become crucial \cite{1.14}.

By construction, the state (\ref{1.18}) is sharply peaked at 
$P^{\k{e}}_j=P^e_j(m), h_{\k{e}}=h_e(m)$ (these values depend explicitly on
$\Sigma,X,P_\gamma,\Pi_\gamma$) and enjoys further important semi-classical
properties as we will describe in detail in the next section. They thus 
are good candidate states to base a semi-classical analysis on.\\
\\
Let us summarize :
\begin{itemize}
\item[i)] {\it Classical Continuum Phase Space}\\
For any given three manifold $\Sigma$
we start from a classical, continuum phase space $({\cal M},\{.,.\})$ 
based on $\Sigma$ with points $m$ and 
a continuum Poisson algebra $\cal O$ of classical observables $O$. 
\item[ii)] {\it Classical Discrete Phase Space}\\
Given a model graph $\k{\gamma}$ we have a classical,
model phase space $({\cal M}_{\k{\gamma}},\{.,.\}_{\k{\gamma}})$ (direct 
sum of copies of $T^\ast SU(2)$, one for each edge of $\k{\gamma}$) with 
points $m_{\k{\gamma}}=\{g_{\k{e}}\}$ and classical 
model Poisson algebra ${\cal O}_{\k{\gamma}}$ of observables 
$O_{\k{\gamma}}$ respectively. No reference to a particular manifold 
$\Sigma$ is made. For any $\Sigma$, 
a subset of ${\cal M}_{\k{\gamma}}$ can be obtained as the image of 
${\cal M}$ based on $\Sigma$ under 
a symplectomorphism $\Phi_{\k{\gamma},\Sigma,X}$ of the form (\ref{1.17}). 
\item[iii)] {\it Discrete Quantum Hilbert Space}\\
The model symplectic manifold is easy to quantize and gives a discrete
quantum theory given by the model graph Hilbert space 
$({\cal H}_{\k{\gamma}},<.,.>_{\k{\gamma}})$
(direct product of of copies of $L_2(SU(2),d\mu_H)$, one for each edge
of $\k{\gamma}$) with coherent states $\psi^t_{\k{\gamma},m_{\k{\gamma}}}$ 
and model quantum commutator algebra $\hat{{\cal O}}_{\k{\gamma}}$ 
of model quantum observables $\hat{O}_{\k{\gamma}}$. By construction, the 
classical limit of 
$({\cal H}_{\k{\gamma}},<.,.>_{\k{\gamma}},\hat{{\cal O}}_{\k{\gamma}})$ 
gives us back
$({\cal M}_{\k{\gamma}},\{.,.\}_{\k{\gamma}}, {\cal O}_{\k{\gamma}})$ which
means that for all $m_{\k{\gamma}},O_{\k{\gamma}}$ the zeroth order term 
in $\hbar$ of the quantities
\ba \label{1.19}
&& <\psi^t_{\k{\gamma},m_{\k{\gamma}}},
\hat{O}_{\k{\gamma}}\psi^t_{\k{\gamma},m_{\k{\gamma}}}>_{\k{\gamma}}
-O_{\k{\gamma}}(m_{\k{\gamma}}) \mbox{ and }
\nonumber\\
&& <\psi^t_{\k{\gamma},m_{\k{\gamma}}},
\frac{[\hat{O}_{\k{\gamma}},\hat{O}'_{\k{\gamma}}]}{it}
\psi^t_{\k{\gamma},m_{\k{\gamma}}}>_{\k{\gamma}}
-\{O_{\k{\gamma}},O'_{\k{\gamma}}\}_{\k{\gamma}}(m_{\k{\gamma}}) 
\ea
and of their fluctuations, e.g. 
$(\Delta O_{\k{\gamma}})_{quant}(m_{\k{\gamma}})$,
vanishes (so-called Ehrenfest property). 
\item[iv)] {\it Continuum Quantum Hilbert Space}\\
Given a model graph $\k{\gamma}$ and an embedding $X$ into some $\Sigma$
we can identify the model Hilbert space ${\cal H}_{\k{\gamma}}$ with 
${\cal H}_\gamma$ where $e=X(\k{e})$ and $h_e:=h_{\k{e}}$. Likewise,
given a model operator $\hat{O}_{\k{\gamma}}$ on ${\cal H}_{\k{\gamma}}$ we 
obtain an operator $\hat{O}_\gamma$ on ${\cal H}_\gamma$ by substituting
multiplication by and differentiation with respect to $h_{\k{e}}$ for
multiplication by and differentiation with respect to $h_e$. 
By this method we obtain diffeomorphism covariant families of operators
$\hat{O}$ and we consider only those that are cylindrically consistent. The 
operators that we have constructed above are of this type and the algebra
of these operators will be called $\hat{{\cal O}}$.

Given $\Sigma,{\cal M}$ and a classical observable $O$ we produce 
by the method described above such a consistently defined family
of operators $\hat{O}=\{\hat{O}_\gamma\}$ based on classical functions
$O_\gamma$ of the graph degrees of freedom. Among all possible graphs 
$\gamma$, by construction there exists at least one
1-parameter family of graphs $\gamma_\epsilon$ such that 
$O(m)=\lim_{\epsilon\to 0} O_{\gamma_\epsilon}(m_{\gamma_\epsilon}(m))$
for any $m\in {\cal M}$. Thus, while for a generic graph $\hat{O}_\gamma$ 
will not have a semi-classical 
interpretation, the operators $\hat{O}_{\gamma_\epsilon}$ do.
The point is now that 
${\cal H}_\gamma\subset{\cal H},<.,.>_\gamma=
<.,.>_{|{\cal H}_\gamma},
\hat{{\cal O}}_\gamma=\hat{{\cal O}}_{|{\cal H}_\gamma}$. 
Therefore, since the continuum limit is coupled to the classical limit
(no separate continuum limit to be taken) we can make the following 
definition :\\
We will say that for given $\Sigma$ the classical limit of 
$({\cal H},<.,.>,\hat{{\cal O}})$ is given by 
$({\cal M},\{.,.\},{\cal O})$ if for all $m,O$ 
the zeroth order of the $\hbar$ expansion of the quantities 
\be \label{1.20}
<\psi_{\gamma_\epsilon,m},
\hat{O}_{\gamma_\epsilon}
\psi_{\gamma_\epsilon,m}>
-O(m) \mbox{ and }
<\psi_{\gamma_\epsilon,m},
\frac{[\hat{O}_{\gamma_\epsilon},
\hat{O}'_{\gamma_\epsilon}]}{it}
\psi_{\gamma_\epsilon,m}>
-\{O,O'\}(m) 
\ee
and of their fluctuations, e.g. $(\Delta O)_{total}(m)$,
vanishes (and that the corrections are in agreement with 
experiment). 
Here $\epsilon=\epsilon(\ell_p,L(m)),a=a(\ell_p,L(m))$ 
or $\epsilon=\epsilon(\ell_p,L_0),a=a(\ell_p,L_0)$ 
take their optimal value as specified above so that they depend only on 
$\ell_p,\Sigma,m$ or $\ell_p,\Sigma,{\cal M}$. Notice that in a label for 
a coherent state we may take $a$ to be $m$-dependent although non-trivial
$m$-dependence of $a$ would spoil the Poisson brackets (\ref{1.11}).
This is because in the model Poisson brackets (\ref{1.12}) we can 
take $a$ to be any function of $m$ without changing them. Alternatively,
$a$ is an $m$-independent parameter to begin with, it just gets fixed
(by optimization) at an $m$-dependent value $a(m)$, thus Poisson brackets 
are to be evaluated by taking $a$ as $m$-independent and after computing
them one should set it equal to $a(m)$.
\end{itemize}
We have thus arrived at a closed diagramme of how to pass from a given 
continuum, classical theory to a continuum, quantum theory and back by 
taking a route through a discrete regime of both theories. Moreover, we have 
derived an explicit criterium for how to decide whether canonical
quantum general relativity has classical general relativity as its
classical limit. The Ehrenfest theorems proved already in \cite{1.3}
and the explicit construction of the operators $\hat{O}_\gamma$
which guarantees that for given $\Sigma,m$ such that 
$\gamma=\gamma_\epsilon,\epsilon=\epsilon(\ell_p,m)$
both its classical and quantum error are small
indicate that this might actually be the case.

A couple of remarks are in order :\\
1.)\\
Our scheme works only if the operators $\hat{O}_\gamma$ leave the 
Hilbert space ${\cal H}_\gamma$ invariant. However, the Diffeomorphism
and Hamiltonian constraint operators that have been constructed in the
literature (e.g. \cite{1.9,1.10}) do not have this property.
Either
one must modify them in such a way that their cylindrical projections 
leave the cylindrical subspaces invariant or we have to modify our scheme
as follows for such operators : If $\hat{O}_\gamma$ maps
${\cal H}_\gamma$ into ${\cal H}_{\gamma'}$ for some  
$\gamma'\not=\gamma$ then find the minimal (in the sense of  
numbers of edges)
$\gamma^{\prime\prime}$ such that 
$\gamma,\gamma'\subset 
\gamma^{\prime\prime}$ and replace 
$\psi_{\gamma_\epsilon,m}$ in (\ref{1.20}) by
$\psi_{\gamma^{\prime\prime}_\epsilon,m}$.
Strictly speaking then, (\ref{1.20}) does not really test the 
continuum operator $\hat{O}$ but only its cylindrical projections
$\hat{O}_\gamma$ which is not what we really want but which is the 
best we can do at the moment. \\
2.)\\
All the coherent states that we defined are labelled by a specific
$\gamma$ and although all of them belong to the continuum Hilbert space
it would be nicer and it would solve the problem just mentioned in 1.),
if we could find a state $\psi_m$ such that its projection onto
${\cal H}_\gamma$ coincides with 
$\psi_{\gamma,m}$. This could be achieved if our family
of states $\psi_{\gamma,m}$ would solve the consistency 
condition that $\psi_{\gamma',m}$ projected onto 
${\cal H}_\gamma$ coincides with $\psi_{\gamma,m}$
for any $\gamma\subset\gamma'$. It is easy to see that
this would require (see also \cite{1.11} for a related observation)
a) to make the parameter $t$ in (\ref{1.14}) a function of the edge
$e$, that is, $t=t(e)$ such that 
$t(e^{-1})=t(e)$ and $t(e\circ e')=t(e)+t(e')$
and b) that $g_{e^{-1}}=g_{e}^{-1}$ and 
$g_{e\circ e'}=g_{e} g_{e'}$ which are precisely the 
defining algebraic relations for a length function and an $SL(2,\Cl)$ 
holonomy respectively. There is no problem 
to generalize our scheme and to make $t$ a function of 
edges. However, our coherent
states are constructed in such a way that the operators
$\hat{P}^{e}_j,\hat{h}_{e}$ are peaked at
$P^{e}_j,h_{e}$ where $g_{e}=
\exp(-iP^{e}_j\tau_j)h_{e}$ and the holonomy property is in conflict
with the interpretation $P^e_j=P^e_j(A,E),h_e=h_e(A)$ given in (\ref{1.10}).
One can, of course, turn the argument the other way around and just 
define a classical complex connection, say $A^\Cl_b(m):=A_b-i e_b/a$ where
$e_b^j$ is the co-triad. Then for given $\Sigma$ and phase 
space point $m$ we can define $g_{e}:=h_e(A^\Cl(m))$ where the 
latter means the holonomy of the complex connection determined by $m$ 
along the embedded edge $e=X(\k{e})$. Then, if $\ell_p\ll\epsilon\ll L_0$ 
the coherent state is peaked at the value 
\be \label{1.21}
h_{e}\approx h_e(A),\; 
P^{e}_j\approx p^e_j(A,E):=
-\frac{1}{2a}\mbox{tr}(\tau_j[\int_e \mbox{Ad}_{h_e(A,x)}(e(x))])
\ee
where $h_e(A,x)$ denotes the holonomy of $A$ along $e$ from the starting
point of $e$ until the point $x\in e$. The advantage of this new map
$\Phi'_{\k{\gamma},\Sigma,X}$ is that we need less structure (no polyhedronal
decomposition, no choices of paths). The disadvantage is that a) (\ref{1.21})
holds only approximately while (\ref{1.18}) holds exactly independent of
the choice of $\epsilon$ and b) there is no closed classical Poisson 
algebra underlying the functions $h_e(A),p^e(A,E)$ : First of all,
the electrical
field ``wants to to be smeared in two rather than in one dimension" so the 
algebra becomes necessarily distributional. Secondly,
since $e_a^j$ or any other Lie algebra valued co-vector 
that can be built from $E^a_j$ is a non-polynomial function of the electric
field, the Poisson $^\ast$ algebra actually does not close. It follows that 
the relation (\ref{1.21})
between the model graph phase space and the classical phase space 
does not survive this interpretation of $g_e$ at the level of 
Poisson brackets. Therefore, the expectation value of the commutator  
between the model elementary graph operators divided by $it$ 
(which is non-distributional) cannot be given by the Poisson brackets of the
functions on the right hand side of (\ref{1.21}) (which is distributional)
as computed with the symplectic structure of $\cal M$.\\
3.)\\
The coherent states that we constructed are {\it kinematical} coherent
states, i.e. they are not annihilated by the constraint operators 
of the theory (in fact, they are not even gauge invariant). One can
rightfully ask how such states can possibly be used in the semiclassical 
analysis. The answer is the following :\\
First of all, one of our aims is find out whether the Hamiltonian 
constraint operators constructed in \cite{1.10} really do have the 
correct classical limit and obviously it is then not allowed to consider
physical coherent states which would be automatically annihilated by them. \\
Secondly, the gauge and diffeomorphism group are unitarily implemented
on the Hilbert space and therefore expectation values and
fluctutations of gauge and spatially diffeomorphism invariant operators 
are in fact gauge and spatially diffeomorphism invariant. This leaves us with 
the issue of observables which are invariant under the motions of the 
Hamiltonian constraint which famously cannot be implemented unitarily.
The problem is, of course, that we do not know any such observable 
explicitly so that it is impossible to test the usefulness of 
kinematical coherent states in full quantum general relativity directly.
However, one can study exactly solvable model systems and it turns out that 
for a large class of such systems \cite{1.12} including those with 
a non-polynomial Hamiltonian constraint and for which the Dirac observables 
are non-polynomial functions of kinematical observables, the expectation
value of these Dirac observables and their commutators divided by $it$
as mesasured by kinematical
coherent states {\it precisely coincides, to zeroth order in $t$, 
with those as measured by 
dynamical coherent states provided we choose $m$ on the constraint surface 
of the phase space, no matter in which gauge}. Moreover, the fluctuations
of these operators are of the same order in $\hbar$. Thus, while
corrections to the classical theory are not independent of one's choice
of kinematical versus dynamical coherent states, one can control 
the size of their difference and the classical limit itself 
seems to be unaffected. This is of course no proof that the same will be 
true in full quantum general relativity but it is a non-trivial consistency
check. Of course, since we only check a finite number of commutators in the
classical limit $\hbar\to 0$ we have only control over the 
infinitesimal dynamics and no a priori control on the finite time evolution
of the system. Fortunately, in principle the finite time evolution is not
necessarily required in order to compute Dirac observables and this is 
all we need since we do not have a Hamiltonian but only a Hamiltonian 
constraint.

Finally, if we consider the GNS representations to be discussed
in section \ref{s4} with respect to a coherent state labelled by 
a point on the constraint surface of the phase space then constraint 
operators 
will have expectation value zero (or at least very close to if not normal 
ordered) although their fluctutations do not vanish. Thus, such 
representations  can be considered as {\it approximately physical 
representations}, a point of view independently reached 
also by Ashtekar (they would be exactly physical if the constraint
operators would be the zero operators).\\
4.)\\
The careful reader may object, as pointed out by Ashtekar and Lewandowski, 
that in order to make our family of operators
cylindrically consistent we must in general introduce the projection 
operators $\hat{P}_\gamma$ which have no classical counterpart and thus
it is actually no longer clear that the classical limit as defined in 
(\ref{1.20}) still gives the desired result. Fortunately, there is no 
problem : At the optimal value of $\epsilon,a$ 
derived in appendix \ref{sb} we find for the projection $\hat{P}_e$
on spin-network states over a single edge $e$ the expectation value 
$1-e^{-\frac{1}{t^{2/3}}}$ where $t\le \ell_p^2/L_0^2$ is a tiny quantity. 
Thus the expectation value is $1+O(t^\infty)$. For a general graph
a similar argument holds provided that the projection operators needed 
in the definition of $\hat{O}_\gamma$ are finite products of the 
$\hat{P}_e$. But this is the case for operators which come from classical
functions defined as integrals over regions, surfaces and curves provided
that the valence of the vertices of $\gamma$ is bounded from above, even
if $\gamma$ has a countably infinite number of edges. It should be clear from
our exposition that on pathological graphs which lack this boundedness
property a good semi-classical behaviour cannot be expected anyway so that 
this is not a restriction.

\section{Coherent States on Cotangent Bundles over Compact Gauge Groups}
\label{s3}

As explained in the previous section, a first step towards constructing 
coherent states for quantum general relativity is to construct such 
states for $SU(2)$. This was achieved in the general setting of compact 
Lie groups by Hall \cite{3.1}. To make our presentation more pedagogic, 
we start by reviewing the situation for the 
harmonic oscillator in one dimension. Here the wave function is given by 
$\psi^t_z (x) = \frac{1}{\sqrt{2\pi t}} e^{-\frac{(z-x)^2}{2t}}$, where 
$z=x_0 - ip_0$ and $t=\frac{\hbar}{m\omega}$, 
and $\psi^t_z $ is an element of $L^2(\Rl,dx)$. As is well-known 
$\psi^t_z (x)$ has a Gaussian shape with peak at $x_0$, or, more 
generally, at $\Re(z)$. Similarly, by transforming  $\psi^t_z (x)$ to 
momentum space one finds that it is peaked there around 
$p_0$, or, again more generally, around $\Im(z)$. Furthermore, the 
sharpness of the peak in each representation is given by $\sqrt{t}$. For 
that reason it is usually stated that coherent states approach the delta 
function as one considers the limit $\hbar$ and therefore $t$ to zero. What 
is really meant by this limit, of 
course, is that the physical quantities characterizing the system under 
consideration - here, for example, $m$ and $\omega$ - are much, much 
larger than the value of $\hbar$. In that sense, the classical limit of, e.g,
expectation values with respect to 
these coherent states is given by the zeroth order and semiclassical 
corrections by the first and higher orders of a power expansion in $t$. 
Further important properties of these states are that they are  
overcomplete, they 
satisfy the Heisenberg uncertainty relation and they are eigenstates of 
the annihilation operator: $(\hat{z} \psi^t_{z_0} )(x) = 
z_0 \psi^t_{z_0} (x)$.

To make the transition from the harmonic oscillator to the compact Lie 
group case, a more technical derivation of the former is helpful. We 
start with the classical phase space $P$, coordinatized by $(x,p)$. By 
choosing a positive function $C$ on $P$ - the
so-called {\it complexifier} - we change to a complex polarization of 
$P$ by defining \begin{eqnarray}
\label{3.1}
z_1:&=& \sum_{n=0}^{\infty} \frac{i^n}{n!} \{x,C\}_n \nonumber\\
z_2:&=& \sum_{n=0}^{\infty} \frac{i^n}{n!} \{p,C\}_n
\end{eqnarray}
where $\{ \cdot,\cdot\}_n$ stands for a multiple Poisson bracket of 
order $n$. The phase space $P$ is now coordinatized by (in general) 
complex coordinates $(z_1,z_2)$. In the case of the harmonic oscillator, 
choosing $C=p^2/2$ results in the usual $z_1=x-ip$ and $z_2=p$. One proceeds
with quantization by replacing Poisson brackets by commutators divided
by $i\hbar$ and by 
promoting classical functions to operators. For the harmonic oscillator 
this can be done without problems: \be
\label{3.2}
\hat{z} = \sum_{n=0}^{\infty} \frac{\hbar^{-n}}{n!} [\hat{x}, \hat{C}]_n 
= \hat{W}_t \hat{x} \hat{W}_t^{-1} 
\ee
with $\hat{W}_t = e^{-\frac{\hat{C}}{t}}$, which for the harmonic 
oscillator is just the heat kernel as then $\hat{C}$ is, up to factors, 
the Laplacian $\Delta$ on $\Rl$. This operator $
\hat{W}_t$ also establishes the connection to the {\it coherent state 
transform} which is a unitary transformation $\hat{U}_t$ from the 
position Hilbert space $L^2(\Rl,dx)$ to the phase space Hilbert 
space (Segal-Bargmann space) $HL^2(\Bbb{C},d\mu)$ of holomorphic 
$L^2$-functions over $\Bbb{C}$. The new measure $\mu$ is 
determined by the unitarity requirement. The transform is defined by 
\be
\label{3.3}
(\hat{U}_t f)(z):= \int dx \rho^t (y,x) f(x) \mid_{y\to z} =:\langle 
\bar{\psi}^t_z ,f\rangle \ee
where $\rho^t(x,y):= (\hat{W}_t \delta_y)(x)$ and $y\to z$ means 
analytic continuation from $y$ to $z$. From these relations one can thus 
deduce the following definition of the harmonic oscillator coherent states:
\be
\label{3.4}
\psi^t_z(x):= (e^{t \Delta /2} \delta_y )(x) \mid_{y\to z}
\ee
which one can check to be identical to the form given at the beginning 
of this section. It should be kept in mind that, while the coherent 
state transform leads to a Hilbert space over $\Bbb{C}$, the associated 
coherent states themselves are still elements 
of the position Hilbert space over $\Bbb{R}$. \\ 
Hall's idea was to follow this construction for compact Lie groups as 
closely as possible. Instead of wavefunctions over $\Bbb{R}$ one 
considers wavefunctions over a compact Lie group $G$, $dx$ is replaced 
by Haar measure $d\mu_H$, the complex label $z=x-ip$ by an element of the 
complexified group $G^{\Bbb{C}}$ that can be 
written in polar decomposition as $g=Hh$ with $h\in G$ and $H$ positiv 
definite hermitian, and finally the Laplacian on $\Bbb{R}$ is replaced 
by the Laplace-Beltrami operator $\Delta_G$ on $G$. Coherent states are then 
analogously defined by
\be
\label{3.10}
\psi^t_g(h):= (e^{t\Delta_G /2} \delta_{\mu_H,h'} )(h)\mid_{h' \to g}.
\ee 
Using the expression for the delta function on Lie groups one arrives at 
the explicit expression \
\be
\label{3.11}
\psi^t_g(h) = \sum_{\pi} d_{\pi} e^{-\frac{t}{2} \lambda_{\pi}} 
\chi_{\pi} (gh^{-1}) \ee
where $\pi$ is (a representant of) an irreducible representation of $G$, 
$d_{\pi}$ its dimension, $\lambda_{\pi}$ the eigenvalue of $\Delta_G$ in 
the representation $\pi$, and $\chi_{\pi}$ denotes the trace in that 
representation. The trace is to be understood in the following sense: 
One first calculates $\chi_{\pi}(h' 
h^{-1})$ with $h'\in G$ and then performs the analytical 
continuation from $h'$ to $g$. 
Hall was also able to show that these coherent states form an overcomplete
system of states on the Hilbert space $L^2(G,d\mu_H)$.

To use Hall's construction for quantum general relativity we specialize
to $G=SU(2)$. Then three additional inputs are required: 
\begin{itemize}

\item One has to establish the connection between the complex group 
element parameterizing the coherent state and the coordinates on the 
classical phase space of general relativity, that is connections and 
electric fields. This is the analogue of $z = x-ip$
in the case of the harmonic oscillator.

\item One has to show that the coherent states so defined have the 
properties important for a (semi-)classical interpretation. Most 
importantly, they should be peaked in the configuration, momentum and 
phase space (Segal-Bargmann) representations around the points in phase space 
specified by $g$, respectively. Also, Ehrenfest theorems should hold, 
that is expectation values of operators and their commutators divided by
$it$ should 
approach the corresponding classical functions and their Poisson 
brackets in the limit $t\to 0$.

\item Finally, one has to extend Hall's framework from single copies of 
the gauge group to multiple ones. In the language of quantum general 
relativity this means the extension from coherent states defined over 
edges to those defined over graphs \cite{1.3}.

\end{itemize}

We begin by defining the complexifier for an edge $e$ by $C_e = 
\frac{1}{2} P^e_i P^e_j\delta^{ij} $ where $P^e_i$ are the momentum 
variables for general relativity as defined in section \ref{s2}. The 
complex holonomies are then given by 
\begin{eqnarray}
\label{3.20}
h_e^{\Bbb{C}}:=g_e &=& \sum_{n=0}^{\infty} \frac{i^n}{n!} 
     \{h_e,C_e\}_n \nonumber \\
  &=& e^{-i\tau_j P^e_j/2} h_e = H_e h_e
\end{eqnarray}
where $H_e$ is positive definite Hermitian and $h_e$ is an element of 
$SU(2)$. Notice that $g_e$ now carries information on the connection as 
well as on the electric field (over the edge $e$). The heat kernel 
operator is $\hat{W}_{t,e}:=e^{\frac{t}{2} \hat
{\Delta}_e}$ with $\hat{\Delta}_e = \frac{1}{4} \delta_{ij} \hat{X}^e_i 
\hat{X}^e_j $, where the $\hat{X}^e_j$ are the derivative operators 
coming from the quantization of the $P^e_j$.
The coherent states for a single edge $e$ have now the form
\be
\label{3.21}
\psi^t_{g_e}(h_e) = \sum_{j=0,\frac{1}{2},1 \ldots} (2j+1) e^{-t j(j+1)/2} 
\chi_j(g_eh_e^{-1})  
\ee
where $g_e$ given by (\ref{3.20}) can be computed, given a point in the 
classical phase space of general relativity as in the previous section 
and $j$ runs over $SU(2)$ spin 
representations. The peakedness or classicality parameter $t$ is given 
by $t=\frac{l^2_p}{a^2}$ where $a$ has dimension of length. Like the mass and 
frequency for the harmonic oscillator it 
characterizes the system under consideration and in practice is fixed
as outlined in section \ref{s2}. Finally we can also 
define the ``annihilation operator''
\be
\label{3.22}
\hat{g}_e:= e^{\frac{t}{2} \hat{\Delta}_e} \hat{h}_e e^{-\frac{t}{2} 
   \hat{\Delta}_e} = e^{\frac{3t}{8}} e^{-i\hat{P}^e_j \tau_j /2} \hat{h}_e
\ee
Here we have used quotation marks because, although the coherent states 
are eigenstates of  $\hat{g}_e$, up to now it is not yet clear whether 
this operator can be used to derive a Fock space structure.
Other properties of the coherent states that follow immediately from 
these definitions are that expectation values of normal ordered 
polynomials of the $\hat{g}_e$ and their adjoints are given by the 
corresponding classical functions (without quantum corrections) and that the 
self-adjoint operators defined by $\hat{x}_e 
:=\frac{1}{2}(\hat{g}^{\dagger}_e + \hat{g}_e)$ and $\hat{y}_e 
:=\frac{1}{2i}(\hat{g}^{\dagger}_e - \hat{g}_e) $ saturate the unquenched
Heisenberg uncertainty relations. Much more work is required to prove the 
peakedness properties so essential for a semiclassical interpretation. 
In the configuration and thus connection representation, e.g., one has 
to prove that 
\be
\label{3.23}
p^t_g(h):= \frac{|\psi^t_g(h)|^2}{\| \psi^t_g \|^2}
\ee
is sharply peaked at $h=h'$ where $g=Hh'$ is the polar decomposition of 
$g$, with sharpness being given by the parameter t. More explicitly, one 
needs to show that an upper bound for $p^t_g(h)$ can be given that 
(roughly) has the form of a Gaussian times 
a quantity that decays to $1$ exponentially fast with $t\to 0$. A first 
look at (\ref{3.21}) reveals that this is not straightforward as the 
convergence producing term $e^{-tj(j+1)/2}$ converges very slowly for 
$t\to 0$. To circumvent this, we employed in the third reference of
\cite{1.3} the so-called Poisson transformation that transforms the sum in 
(\ref{3.21}) into a sum where the exponential contains $t$ in the 
denominator. This speeds up convergence dramatically and thus allows one 
to derive the desired results, even though calculations are still very 
tedious to perform. 

By similar methods peakedness was shown in the momentum and phase space 
representation. In addition, as a further check-up and for illustration 
purposes, these calculations were also performed numerically with a 
confirmation of the analytical results, see
\cite{1.3} for some graphics. As a result, the coherent states defined 
above have been shown to be peaked sharply around a point in the 
classical phase space (in contrast, e.g., to the so-called weaves 
\cite{1.5a}, which so-far had been used to approximate classical spacetimes, 
but were peaked only in the momentum 
configuration). They thus serve as a good starting point to approximate 
classical spacetimes (remember that quantum states can never reproduce a 
classical state exactly, due to the uncertainty relations), even more so as 
we were able to derive Ehrenfest theorems for them in the fourth
reference of \cite{1.3}. 
This means that expectation values of arbitrary polynomials in the 
$\hat{h}_e$ and $\hat{p}^j_e$ with respect to the coherent states 
approach the corresponding polynomial of $h_e$ and $p^j_e$ on phase space in 
the limit $t\to 0$. Also, expectation values of commutators divided by 
$it$ approach the value of the corresponding Poisson bracket, which 
means that with respect to the coherent states, infinitesimally, the quantum 
dynamics reproduces the classical dynamics.   

Finally, coherent states over graphs are constructed by simply tensoring 
together those over the respective edges, that is \be
\label{3.27}
\psi^t_{\{g\},\gamma}(\{h\}) := \prod_{e\in E(\gamma)} \psi^t_{g_e}(h_e)
\ee
where $E(\gamma)$ is the set of all edges of the graph $\gamma$ and 
$\{g\},\{h\}$ denote the collections $g_{e_1} \ldots g_{e_n}$,$h_{e_1} 
\ldots h_{e_n}$ of the associated group elements, respectively. The 
peakedness and Ehrenfest properties discussed above for the single-edge case 
also go through for the case of graphs. This follows from the fact that 
peakedness on single edges implies peakedness on the whole graph and 
vice versa, and from the fact that operators on different edges commute.
This tensor product structure of the coherent states brings us naturally to
the next subject.

\section{The Infinite Tensor Product Extension}
\label{s4}

Quantum field theory on curved spacetimes is best understood if the 
spacetime is actually flat Minkowski space on the manifold $M=\Rl^4$.
Thus, when one wants to compute the low energy limit of canonical quantum
general relativity to show that one gets the standard model (plus 
corrections) on a background metric one should do this first for the 
Minkowski background metric. Any classical metric is macroscopically
non-degenerate. Since in the Hilbert space that has 
been constructed for loop quantum gravity (see \cite{1.1} for a review and 
\cite{4.1} for more details) the quantum excitations of the 
gravitational field are concentrated on the edges of a
graph, in order that, say, the expection values of the volume operator 
for any macroscopic region is non-vanishing and changes smoothly as
we vary the region, the graph must fill the 
initial value data slice densely enough, the mean separation between
vertices of the graph must be much smaller than the size of the region
(everything is measured by the three metric, determined by the four metric
to be approximated, in this case the Euclidean one). Now $\Rl^4$ is 
spatially non-compact and therefore such a graph 
must necessarily have an at least {\it countably infinite} 
number of edges whose union has {\it non-compact} range.

The Hilbert spaces in use for loop quantum gravity have a dense 
subspace consisting of finite linear combinations of so-called
cylindrical functions labelled either by a piecewise analytic graph with
a {\it finite} number of edges or by a so-called web, a piecewise smooth 
graph determined by the union of a {\it finite} number of smooth
curves that intersect in a controlled way \cite{4.2}, albeit possibly
a countably infinite number of times in accumulation points of edges and
vertices. Moreover, in both cases the 
edges or curves respectively are contained in compact subsets of the 
initial data hypersurface. These categories of graphs will be denoted by
$\Gamma^\omega_0$ and $\Gamma^\infty_0$ respectively
where $\omega,\infty,0$ stands for analytic, smooth and compactly supported
respectively. Thus, the only way that the current Hilbert
spaces can actually produce states depending on a countably infinite graph
of non-compact range is by choosing elements in the closure of these
spaces, that is, states that are countably infinite linear combinations 
of cylindrical functions.

The question is whether it is possible to produce semi-classical states of 
this form, that is, $\psi=\sum_n z_n \psi_{\gamma_n}$ where 
$\gamma_n$ is either a finite piecewise analytic graph or a web, $z_n$ is 
a complex number and we are summing over the integers. It is easy to see 
that this is not the case : Minkowski
space has the Poincar\'e group as its symmetry group and thus we will have
to construct a state which is at least invariant under 
(discrete) spatial translations. This forces the $\gamma_n$ to be 
translations of $\gamma_0$ and $z_n=z_0$. Moreover, the dependence 
of the state on each of the edges has to be the same and therefore 
the $\gamma_n$ have to be mutually disjoint. It follows that the norm
of the state is given by
\be \label{4.1}
||\psi||^2=|z|^2([\sum_n 1][1-|<1,\psi_{\gamma_0}>|^2]+
[\sum_{n} 1]^2 |<1,\psi_{\gamma_0}>|^2)
\ee
where we assumed without loss of generality that $||\psi_{\gamma_0}||=1$
and we used the diffeomorphism invariance of the measure and $1$ is the 
normalized constant state. By the Schwarz inequality the first term is 
non-negative and convergent only if $\psi_{\gamma_0}=1$ while the second
is non-negative and convergent only if $<1,\psi_{\gamma_0}>=0$. Thus
the norm diverges unless $z=0$.

This caveat is the source of its removal : We notice that the formal state 
$\psi:=\prod_n \psi_{\gamma_n}$
really depends on an infinite graph and has unit norm if we formally
compute it by $\lim_{N\to\infty} ||\prod_{n=-N}^N \psi_{\gamma_n}||=1$
using disjointness of the $\gamma_n$. The only problem is that this
state is not any longer in our Hilbert space, it is not the Cauchy 
limit of any state in the Hilbert space : Defining $\psi_N:=
\prod_{n=-N}^N \psi_{\gamma_n}$ we find 
$|<\psi_N,\psi_M>|=|<1,\psi_{\gamma_0}>|^{2|N-M|}$ so that $\psi_N$ is 
not a Cauchy sequence unless $\psi_{\gamma_0}=1$. However, it turns out 
that it belongs to the {\it Infinite Tensor Product (ITP) extension} of the 
Hilbert space.

To construct this much larger Hilbert space we must first describe
the class of embedded graphs that we want to consider. We will consider 
graphs
of the category $\Gamma^\omega_\sigma$ where $\sigma$ now stands for
countably infinite. More precisely, an element of 
$\Gamma^\omega_\sigma$ is the union of a countably infinite  
number of analytic, 
mutually disjoint (except possibly for their endpoints) curves called 
edges of compact or non-compact range which have no accumulation points of
edges or vertices. In other words, the restriction of the graph 
to any compact subset of the hypersurface looks like an element of 
$\Gamma^\omega_0$. These are precisely the kinds of graphs that one 
would consider in the thermodynamic limit of lattice gauge theories 
and are therefore best suited for our semi-classical considerations since 
it will be on such graphs that one can write actions, Hamiltonians and the 
like. 

The construction of the ITP of Hilbert spaces is due to von Neumann
\cite{4.3} and already more than sixty years old. We will try to outline
briefly some of the mahematical notions involved (see the fifth reference in
\cite{1.3} for a concise summary of the most important  
definitions and theorems). \\
\\
Let for the time being $I$ be any
index set whose cardinality $|I|=\aleph$ takes values in the set
of non-standard numbers (Cantor's alephs). Suppose that for each 
$e\in I$ we have a Hilbert space ${\cal H}_e$ with scalar product
$<.,.>_e$ and norm $||.||_e$. 
For complex numbers $z_e$ we say that $\prod_{e\in I} z_e$
{\it converges} to the number $z$ 
provided that for each positive number $\delta>0$ there 
exists a finite set $I_0(\delta)\subset I$ such that for any other finite
$J$ with $I_0(\delta)\subset J\subset I$ it holds that 
$|\prod_{e\in J} z_e -z|<\delta$. We say that $\prod_{e\in I} z_e$
is {\it quasi-convergent} if $\prod_{e\in I} |z_e|$ converges. 
If $\prod_{e\in I} z_e$ is quasi-convergent but not convergent 
we define $\prod_{e\in I} z_e:=0$. Next we say that for 
$f_e\in {\cal H}_e$ the ITP $\otimes_f:=\otimes_e f_e$ is a $C_0$ vector
(and $f=(f_e)$ a $C_0$ sequence) if 
$||\otimes_f||:=\prod_{e\in I} ||f_e||_e$ converges to
a non-vanishing number. Two $C_0$ sequences $f,f'$ are said to be 
strongly resp. weakly equivalent provided that
\be \label{4.2}
\sum_e |<f_e,f'_e>_e-1| \mbox{ resp. } \sum_e ||<f_e,f'_e>_e|-1| 
\ee
converges. The strong and weak equivalence class of $f$ is denoted by 
$[f]$ and $(f)$ respectively and the set of strong and weak equivalence
classes by $\cal S$ and $\cal W$ respectively. We define the ITP
Hilbert space ${\cal H}^\otimes:=\otimes_e {\cal H}_e$ to be the closed 
linear span of all $C_0$ vectors. Likewise we define
${\cal H}^\otimes_{[f]}$ or ${\cal H}^\otimes_{(f)}$ to be the closed
linear spans of only those $C_0$ vectors which lie in the same strong 
or weak equivalence class as $f$. The importance of these notions is that
the determine much of the structutre of ${\cal H}^\otimes$, namely : \\
1) All the ${\cal H}^\otimes_{[f]}$ are isomorphic and mutually orthogonal.\\
2) Every ${\cal H}^\otimes_{(f)}$ is the closed direct sum of all the 
${\cal H}^\otimes_{[f']}$ with $[f']\in {\cal S}\cap (f)$.\\ 
3) The ITP ${\cal H}^\otimes$ is the closed direct sum of all the 
${\cal H}^\otimes_{(f)}$ with $(f)\in {\cal W}$.\\ 
4) Every ${\cal H}^\otimes_{[f]}$ has an explicitly known orthonormal 
von Neumann basis.\\
5) If $s,s'$ are two different strong equivalence classes in the same 
weak one then there exists a unitary operator on ${\cal H}^\otimes$
that maps ${\cal H}^\otimes_s$ to ${\cal H}^\otimes_{s'}$, otherwise 
such an operator does not exist, the two Hilbert spaces are unitarily
inequivalent subspaces of ${\cal H}^\otimes$.\\ 
Notice that two isomorphic Hilbert spaces can always be mapped into each 
other such that scalar products are preserved (just map some orthonormal 
bases) but here the question is whether this map can be extended 
unitarily to all of
${\cal H}^\otimes$. Intuitively then, strong classes within the same weak 
classes describe the same physics, those in different weak classes describe
different physics such as an infinite difference in energy, magnetization,
volume etc. See \cite{4.4} and references therein as well as 
appendix \ref{sc} for illustrative examples.

Next, given a bounded operator 
$a_e$ on ${\cal H}_e$ (notice that closed unbounded operators have a polar  
decomposition into an unitary and a self-adjoint piece and that a 
self-adjoint operator is completely determined by its bounded spectral 
projections so that restriction to bounded operators is no loss of
generality) we can extend it in the natural way to 
${\cal H}^\otimes$ by defining $\hat{a}_e$ densely on $C_0$ vectors through
$\hat{a}_e \otimes_f=\otimes_{f'}$ with $f'_{e'}=f_{e'}$ for $e'\not=e$
and $f'_e= a_e f_e$. It turns out that the algebra of these extended 
operators for a given label is automatically a von Neumann algebra for 
${\cal H}^\otimes$ and we will call the weak closure of all these algebras 
the von Neumann algebra ${\cal R}^\otimes$ of local operators.

Given these notions, the strong equivalence class Hilbert spaces can be 
characterized 
further as follows. First of all, for each $s\in {\cal S}$ one can 
find a representant $\Omega^s\in s$ such that $||\Omega^s||=1$. Moreover, 
one can show
that ${\cal H}^\otimes_s$ is the closed linear span of those $C_0$ vectors
$\otimes_{f'}$ such that $f'_e=\Omega^s_e$ for all but finitely
many $e$. In other words, the strong
equivalence class Hilbert spaces are irreducible subspaces for 
${\cal R}^\otimes$, $\Omega^s$ is a cyclic vector for ${\cal H}^\otimes_s$
on which the local operators annihilate and create local excitations
and thus, if $I$ is countable, ${\cal H}^\otimes_s$ is actually separable.
We see that we make naturally contact with Fock space structures,
von Neumann algebras and their factor type classification \cite{4.5}
(modular theory) and algebraic quantum field theory \cite{4.6}. 
The algebra of operators on the ITP which are not local do not have 
an immediate interpretation but it is challenging that they 
map between different weak equivalence classes and thus change the physics
in a drastic way. It is speculative that this might in fact enable us to 
incorporate dynamical topology change in canonical quantum gravity.

A number of warnings are in order :\\
1) Scalar multiplication is not multi-linear ! That is, if $f$ and 
$z\cdot f$ are $C_0$ sequences where $(z\cdot f)_e=z_e f_e$ for some
complex numbers $z_e$ then $\otimes_f=(\prod_e z_e)\;\otimes_f$ is in general
wrong, it is true if and only if $\prod_e z_e$ converges.\\
2) Unrestricted use of the associative law of tensor products is false !
Let us subdivide the index set $I$ into mutually disjoint
index sets $I=\cup_\alpha I_\alpha$ where $\alpha$ runs over some other index
set $A$. One can now form the different ITP ${\cal H}^{\prime\otimes}
=\otimes_\alpha {\cal H}^\otimes_\alpha,\;
{\cal H}^\otimes_\alpha=\otimes_{e\in I_\alpha} {\cal H}_e$.
Unless the index set $A$ is finite, a generic $C_0$ vector of 
${\cal H}^{\prime\otimes}$ is orthogonal to all of
${\cal H}^\otimes$. This fact has implications for quantum gravity 
which we outline below.\\
\\
After this mathematical digression we 
now come back to canonical quantum general relativity. In applying 
the above concepts we arrive at the following surprises :
\begin{itemize}
\item[i)] First of all, we 
fix an element $\gamma\in \Gamma^\omega_\sigma$
and choose the countably infinite index set $I:=E(\gamma)$, the edge
set of $\gamma$. If $|E(\gamma)|$ is finite then the ITP Hilbert space
${\cal H}^\otimes_\gamma:=\otimes_{e\in E(\gamma)} {\cal H}_e$ is naturally
isomorphic with the subspace ${\cal H}^{AL}_\gamma$ of the Ashtekar 
Lewandowski
Hilbert space obtained as the closed linear span of cylinder functions
over $\gamma$. However, if $|E(\gamma)|$ is truly infinite then 
a generic $C_0$ vector of ${\cal H}^\otimes_\gamma$ is orthogonal 
to any possible ${\cal H}^{AL}_{\gamma'}$, $\gamma'\in \Gamma^\omega_0$.
Thus, even if we fix only one $\gamma\in \Gamma^\omega_\sigma$, the total
${\cal H}^{AL}$ is orthogonal to almost every element of ${\cal 
H}^\otimes_\gamma$.
\item[ii)] Does ${\cal H}^\otimes_\gamma$ have a measure theoretic 
interpretation as an $L_2$ space ? By the Kolmogorov theorem \cite{4.7}
the infinite product of probability measures is well defined and thus one
is tempted to identify ${\cal H}^\otimes_\gamma=\otimes_e L_2(SU(2),d\mu_H)$
with ${\cal H}^{AL\prime}_\gamma:=L_2(\times_e SU(2),\otimes_e d\mu_H)$. 
However, this cannot be 
the case, the ITP Hilbert space is non-separable (as soon as $\dim({\cal 
H}_e)>1$ for 
almost all $e$ and $|E(\gamma)|=\infty$) while the latter Hilbert space
is separable, in fact, it is the subspace of ${\cal H}^{AL}$ consisting
of the closed linear span of cylindrical functions over $\gamma'$
with $\gamma'\in \Gamma^\omega_0\cap E(\gamma)$.
\item[iii)] Yet, there is a relation between ${\cal H}^\otimes_\gamma$ 
and ${\cal H}^{AL}$ through a construction called the inductive limit
of Hilbert spaces : We can find a directed sequence of elements $\gamma_n\in 
\Gamma^\omega_0\cap E(\gamma)$, that is, $\gamma_m\subset\gamma_n$ for 
$m\le n$, such that $\gamma$ is its limit in $\Gamma^\omega_\sigma$.
The subspaces ${\cal H}^{AL}_{\gamma_n}\subset{\cal H}^{AL}$ are isometric 
isomorphic with the subspaces of ${\cal H}^\otimes_\gamma$ given by the 
closed linear span of vectors of the form $\psi_{\gamma_n}\otimes[
\otimes_{e\in E(\gamma-\gamma_n)} 1]$ where $\psi_{\gamma_n}\in
{\cal H}^{AL}_{\gamma_n}\equiv {\cal H}^\otimes_{\gamma_n}$ which provides 
the 
necessary isometric monomorphism to display ${\cal H}^\otimes_\gamma$
as the inductive limit of the ${\cal H}^{AL}_{\gamma_n}$.
\item[vi)] So far we have looked only at a specific 
$\gamma\in\Gamma^\omega_\sigma$. We now construct the total Hilbert space
\be \label{4.3}
{\cal H}^\otimes:=\overline{\cup_{\gamma\in \Gamma^\omega_\sigma}
{\cal H}^\otimes_\gamma}
\ee
equipped with the natural scalar product derived in the fifth reference
of \cite{1.3}.
This is to be compared with the 
Ashtekar-Lewandowski Hilbert space 
\be \label{4.4}
{\cal H}^{AL}:=
\overline{\cup_{\gamma\in \Gamma^\omega_0}{\cal H}^{AL}_\gamma}
=\overline{\cup_{\gamma\in \Gamma^\omega_\sigma}{\cal H}^{AL\prime}_\gamma}
\ee
The identity in the last line enables us to specify the precise sense in 
which ${\cal H}^{AL}\subset{\cal H}^\otimes$ : For any 
$\gamma\in\Gamma^\omega_\sigma$ the space
${\cal H}^{AL\prime}_\gamma$ is isometric isomorphic as specified in 
iii) with the strong eqivalence class Hilbert subspace 
${\cal H}^\otimes_{\gamma,[1]}$ where $1_e=1$ is the constant function 
equal to one. Thus, the the Ashtekar Lewandowski Hilbert space 
describes the local excitations of the Ashtekar-Lewandowski ``vacuum''
$\Omega^{AL}$ with $\Omega^{AL}_e=1$ for any possible analytic path $e$.

Notice that both Hilbert spaces are non-separable, but there are two
sources of non-separability : the Ashtekar Lewandowski Hilbert space is
non-separable because $\Gamma^\omega_0$ has uncountable infinite
cardinality. This is also true for the ITP Hilbert since 
$\Gamma^\omega_0\subset\Gamma^\omega_\sigma$ but
it has an additional character of non-separability : even for fixed $\gamma$
with an infinite number of edges the Hilbert space
${\cal H}^\otimes_\gamma$ splits into an uncountably infinite number of 
mutually orthogonal strong equivalence class Hilbert spaces and 
${\cal H}^{AL\prime}_\gamma$ is only one of them.
\item[v)] Recall that spin-network states \cite{1.7} form a basis for 
${\cal H}^{AL}$. The result of iv) states that they are no longer
a basis for the ITP. The spin-network basis is in fact the von Neumann
basis for the strong equivalence class Hilbert space determined 
by $[\Omega^{AL}]$ but for the others we need uncountably infinitely
many other bases, even for fixed $\gamma$. The technical reason for this is 
that,
as remarked above, the unrestricted associativity law fails on the ITP.
\end{itemize}
We would now like to justify this huge blow up of the original Ashtekar
Lewandowski Hilbert space from the point of view of physics. Clearly,
there is a blow up only when the initial data hypersurface is non-compact
as otherwise $\Gamma^\omega_0=\Gamma^\omega_\sigma$.
\begin{itemize}
\item[a)] Let us fix a three manifold $\Sigma$
and a graph $\gamma\in\Gamma^\omega_\sigma$ in order to describe 
semi-classical physics on that graph as outlined in section \ref{s2}.
Given a classical initial data 
set $m$ we can construct a coherent state $\psi_{\k{\gamma},\Sigma,X,m}$ 
which in fact is a $C_0$ vector $\otimes^\gamma_{f_m}$ for 
${\cal H}^\otimes_{\k{\gamma}}$ of unit norm. 
This coherent state can be considered as a ``vacuum''
or ``background state'' for quantum field theory on the associated
spacetime.
As remarked above, the corresponding strong eqivalence class Hilbert space
${\cal H}^\otimes_{\gamma,[f_m]}$ is obtained by acting on the 
``vacuum'' by local operators (where local means that
only finitely many edges are affected), resulting in a space isomorphic with 
the familar Fock spaces and which is separable. In this sense, the
fact that ${\cal H}^\otimes_\gamma$ is non-separable, being an 
uncountably infinite direct
sum of strong equivalence class Hilbert spaces, simply accounts for the 
fact that in quantum gravity {\it all vacua have to be considered 
simultaneously, there is no distinguished vauum as we otherwise would
introduce a backgrond dependence into the theory}.
\item[b)] The Fock space structure of the strong equivalence classes
immediately suggests to try to identify suitable excitations of 
$\psi_{\gamma,m}$ as graviton states propagating on a spacetime
fluctuating around the classical background determined by $m$
\cite{4.8}. \\
Also, it is easy to check whether for different solutions
of Einstein's equations 
the associated strong equivalence classes lie in different
weak classes and are thus physically different. For instance,
preliminary investigations indicate that Schwarzschild black hole spacetimes 
with different masses lie in the {\it same} weak class. Thus, {\it unitary}
black hole evaporation and formation seems not to be excluded from the 
outset.
\item[c)] From the point of view of ${\cal H}^{AL\prime}_\gamma$
the Minkowski coherent state is an everywhere excited state like a 
thermal state, the strong classes $[\Omega^{AL}]$ and $[f_m]$ for
Minkowski data $m$ are orthogonal and lie in different weak classes.
The state $\Omega^{AL}$ has no obvious semi-classical interpretation 
in terms of coherent states for any classical spacetime.
\item[d)] If we look at the construction in the fourth reference of 
\cite{1.1} then, provided one 
looks {\it only at local} operators in the {\it bulk} of a non-compact three 
manifold, we can avoid the ITP by using the GNS construction. 
In fact, specifying a GNS state is the same as looking at a specific
representant of a strong equivalence class of our ITP Hilbert space,
in that sense the construction in the fourth reference in \cite{1.1} gives 
only a special subspace of our ITP. More importantly, however, our 
framework is much more general : First of all, we can incorporate non-local 
operators such as Hamiltonian 
constraint operators smeared against test fields of rapid decrease rather 
than compact support (as required in order to discuss supertranslations)
or matter Hamiltonians coupled to gravity important in order to 
obtain effective matter Hamiltonians of the standard model in a 
gravitational state fluctuating around Minkowski space. Secondly, we can
treat operators that
are defined not only in the bulk but also at spatial infinity thus enabling
us to consider electromagnetic charges (when coupling matter) and the 
Poincar\'e charges. Moreover, our approach provides a unifying and 
rigorous framework of how to address the infinite volume (thermodynamic)
limit of the theory.
\item[e)] Another important kind of operators that is crucial in 
order to make contact with perturbative quantum gravity and string 
theory is the graviton annihilation and creation operator labelled
by a specific momentum (and helicity) mode and which is therefore a
maximally non-local operator. Once we 
have identified suitable graviton Fock states propagating
on a fluctuating quantum metric as defined through the excitations of 
a unit norm representant of the corresponding strong equivalence class,
we should be able, in principle to make contact with perturbative quantum 
field theory and thus to deliver significant evidence that loop quantum
gravity is not just some completely spurious sector of quantum gravity.
\end{itemize}

\section{Applications and Outlook}
\label{s5}

Given the tools provided in \cite{1.3,1.5} one can start investigating many 
fascinating physical problems starting from non-perturbative quantum 
general relativity. We will list a few of them.
\begin{itemize}
\item[i)] {\it Generalization to Higher Ranks}\\
In \cite{1.3} peakedness and Ehrenfest theorems were analytically proved
for rank one gauge groups and a generalization of the proof to higher ranks
was sketched. It would be important to fill in the details at least
for $G=SU(3)$, if not for arbitrary compact gauge groups.
\item[ii)] {\it Quantum Field Theory on Curved Spacetimes}\\
As already mentioned in the main text, choosing a coherent state peaked
on the initial values of a solution to Einstein's equations one obtains
a cyclic state for a Fock-like representation. It would be important to 
isolate suitable graviton states in such a representation in order to make
contact with perturbative quantum gravity and string theory. The same 
can be done when coupling matter and one could, for instance, study 
photons propagating on fluctuating quantum spacetimes.
\item[iii)] {\it Renormalization Group and Diffeomorphism Group}\\
Given a Fock like structure one can try to set up an analog of the 
conventional renormalization techniques. One expects that from 
non-perturbative quantum gravity no UV singularities arise (see \cite{1.10})
for hint in that direction). But then one would like to see how they get
absorbed as compared to the conventional perturbative approaches and 
the obvious guess is that it is the diffeomorphism invariance of the theory
that lets us get rid of them (since there is no background metric, one cannot
tell from a fundamental point of view whether a momentum gets large or 
small). It should be possible to nail down the precise mechanism for 
UV-finiteness with our tools. 
\item[iv)] {\it Avoidance of Classical Singularity Theorems}\\
One can plug into the label of a coherent state a solution $m(\tau)$
of Einstein's equations (in some gauge) which becomes singular after 
some time lapse $\tau_0$. It would be challenging to see whether the 
classically singular function remains finite in the sense of expectation 
values of the corresponding operator at $\tau=\tau_0$. The recent 
regularity findings obtained within loop quantum cosmology \cite{5.1} which 
uses the techniques developed in \cite{1.10} are an indication that this
actually could be the case in the full theory as well.
\item[v)] {\it Corrections to the Standard Model}\\
One of the major motivations for the construction of coherent states 
is, of course, the question whether our non-perturbative quantum theory
has the correct classical limit. But beyond that one can start computing
corrections to the standard model, a development that has just 
started. See for instance the two first references in \cite{1.1} discussing
Poincar\'e invariance violating effects that are not yet ruled out to
lie in the detectable regime. In order to improve these calculations it 
would be desirable to have a fast diagonalization computer code for the 
volume operator \cite{1.4,5.2} since it plays an important role in the 
quantization of the Hamiltonian (constraints) \cite{1.10}.
\end{itemize}
~~~~~~~~~~~~~~~~~~~~~~~~~~~~~~~~~~~~~~~~\\
\\
\\
{\large Acknowledgements}\\
\\
We thank Alexandro Corichi, Rodolfo Gambini, Brian Hall, 
Bei-Lok Hu, Ted Jacobson,  
Hugo Morales-Tecotl and, especially, Luca Bombelli and Jurek Lewandowski 
as well as the members of the quantum gravity division 
of the Center for Gravitational Physics and Geometry Martin Bojowald,
Olaf Dreyer, Stephen Fairhurst, Amit Gosh, Badri Krishnan, Jacek Wisniewski  
and, especially, Abhay Ashtekar and Jorge Pullin for very productive 
discussions about semiclassical states, which strongly clarified our 
understanding of the appearance of different 
scales in quantum gravity, the difference between coherent and  
semi-classical states and the difference between kinematical and 
dynamical semi-classical states. Their inputs and criticisms 
influenced our point of view and lead us to 
our construction of approximate operators and our fluctuation optimization 
procedure. \\
We also thank Brian Hall for clarifying 
discussions about infinite tensor products.\\
H. Sahlmann and O. Winkler thank the Studienstiftung des Deutschen 
Volkes for financial support.

\begin{appendix}

\section{Construction of Approximate Operators} 
\label{sa}

In this section we construct an approximate area operator for a specific
class of surfaces.\\

Suppose we are given some embedded graph $\gamma=X(\k{\gamma})$ (here given
as the image under the embedding $X$ of a model graph $\k{\gamma}$)
and look at 
the area function $A_S(m):=\int_{Y^{-1}(S)} d^2y\sqrt{\det(Y^\ast q)}$
of an embedded surface $S$ (here given as the image under the embedding 
$Y$ of a model two surface $\k{S}$ into $\Sigma$) where $q$ denotes the 
three metric determined by $m$. We choose a polyhedronal 
decomposition $P_\gamma$ of $\Sigma$ dual to $\gamma$ and a corresponding
system of paths $\Pi_\gamma$ inside its faces. Our first task is 
to write down a function of the correspeonding graph degrees of freedom
$g_e=(h_e,P^e_j)$ that approximates $A_S$. As compared to the construction
in the main text we choose here for simplcity $X_0=X,\varphi_{\gamma,S}=
\mbox{id}_\Sigma$.

Without specifying exactly
the topology of $\k{\gamma},\k{S}$ it is difficult to write down an 
explicit formula, so let us assume for simplcity that $\k{\gamma}$
has the topology of a (possibly infinite) cubic lattice. In this case
model edges $\k{e}_I(\k{v})$ can be labelled by a direction index 
$I=1,2,3$ and a vertex index $\k{v}\in\Zl^3$ and we think of them 
as given by unit intervals along the coordinate axes of $\Rl^3$.
Furthermore, we take as $P_\gamma$ the image under $X$ of the faces 
$S^I(v)$ (dual to $e_I(v)$ of the model graph given by the translation 
of $\k{\gamma}$
by the vector $(1/2,1/2,1/2)$ in $\Rl^3$ and for $\Pi_\gamma$
we take the image under $X$ of the natural radial paths in each of the faces.
Likewise, for definiteness, we assume that $\k{S}$ has the topology of a 
square so that
$u,v$ range over the interval $(-N,N)$ where $N$ is an integer. 
The embedding $X:\;\Rl^3\mapsto \Sigma; \vec{t}\mapsto
X(\vec{t})$ provides a (local) coordinate chart so that we can think of 
$Y$ as a composition 
$Y^a(u,v)=X^a(\vec{t}(u,v))$. The coordinates $\vec{t},u,v$ are taken to 
be dimensionless. We can then write the area function explicitly as 
\be \label{a.1}
A_S(q)=\int_{[-N,N]^2}du dv
\sqrt{
[(E^a_j(X(\vec{t}) n_a^I(\vec{t}))_{\vec{t}=\vec{t}(u,v)} n_I(u,v)]^2}
\ee
where $n_a^I(\vec{t})=\frac{1}{2} \epsilon_{abc} \epsilon^{IJK}
X^b_{,J}(\vec{t}) X^c_{,J}(\vec{t})$ and 
$n_I(u,v)=\epsilon_{IJK} t^J_{,u}(u,v) t^K_{,v}(u,v)$.

Let us now partition $\k{S}$ into the maximal, connected, open pieces 
$\k{S}_k$ which 
lie completely within a coordinate cube of $\Rl^3$, let $\k{v}_k$ be 
the vertex of that cube, choose any $(u_k,v_k)\in \k{S}_k$ and let 
$\mu_k:=\int_{\k{S}_k} du dv$. Then, given any point $m\in {\cal M}$ 
we denote by $P^I_j(v,m):=P^{e_I(v)}_j(m)$ and $n_I(k):=n_I(u_k,v_k)$. 
With these choices we arrive at the gauge invariant approximate function
\be \label{a.2}
A_{S,\gamma}(m):=a^2\sum_k \mu_k
\sqrt{[P^I_j(v_k,m) n_I(k)]^2}
\ee
where $v_k=X(\k{v}_k)$ which now depends explicitly on our choices 
(collectively denoted 
by $\gamma$) as well as the connection and so is a function of $m$
and not only of $q_{ab}$. By construction, for fixed $S$ (\ref{a.2}) 
is a Riemann sum for the integral (\ref{a.1}) and thus converges to 
it pointwise in $\cal M$ in the limit that the sum over $k$ involves 
infinitely many terms
(the model lattice becomes finer and finer within the model surface).
Notice that the coefficients $\mu_k,n_I(k)$ and the sum over 
the $\k{S}_k$ do not depend on 
$X$ or $m$ and thus (\ref{a.2}) defines a well-defined
function on $\cal M$. Moreover, (\ref{a.2}) is
diffeomorphism covariant since the only dependence on $X$ rests in the 
$P^I(v,m)$ while the $\mu_k,n_I(k),\k{S}_k$ are diffeomorphism invariant.
Equation (\ref{a.2}) thus serves as a suitable departure point for 
quantization.

According to our general programme we define now the substitute operator 
\be \label{a.3}
\hat{A}_{S,\gamma}=a^2\sum_k \mu_k
\sqrt{[\hat{P}^I_j(v_k) n_I(k)]^2}
\ee
which is to be thought of as an operator on ${\cal H}_\gamma$ since it is 
already consistently defined on every subgraph of $\gamma$. 
By doing a similar construction for an arbitrary graph
and by enforcing cylindrical consistency we in principle (we do not 
know it explicitly but must construct it graph by graph) arrive 
at a ``continuum'' operator $\hat{A}_S$.

Let us now consider in a specific example for $\gamma,S$ how the expectation
values are computed. Consider the embeddings into $\Sigma:=\Rl^3$
\be \label{a.4}
X^a(\vec{t}):=\epsilon \delta^a_I t^I,\;Y^a(u,v)=\epsilon(u,v,\tau u),\;
\vec{t}(u,v)= (u,v,\tau u)
\ee
where $0\le\tau\le 1$ is a fixed parameter and $u,v\in[-N,N]$. Equation 
(\ref{a.4}) defines a plane parallel to the $t^2$ direction and 
which is tilted against the $t^1$ direction by an angle $\alpha$ defined
by $\tan(\alpha)=\tau$ (for $\alpha\ge \pi/4$ interchange the directions
of $t^1,t^2$). We will not consider the most general case for which it is 
difficult to write down an explicit formula but rather consider the
case that $\tau=1/M$ where $M\ge$ is an integer such that $N/M$ is also
a natural number. In this case we are looking at the vertices 
$\k{v}(n_1,n_2)=(n^I)=(n_1,n_2,n_3(n_1)),\;n_1,n_2=-N,-N+1,..,N-1$ 
where 
$n_3(n_1)=-N\tau+[(n_1+N)\tau]$ (Gauss bracket). The $\k{S}_k$ are simply
model unit squares so that $\mu_k=1$. We easily compute $(n_I(k))=
(-\tau,0,1)$ so that (\ref{a.3}) becomes 
\be \label{a.5}
\hat{A}_{S,\gamma}=a^2\sum_{n_1,n_2}
\sqrt{[\hat{P}^3_j(v(n_1,n_2)-\tau\hat{P}^1_j(v(n_1,n_2)]^2}
\ee
where $v(n_1,n_2)=X(\k{v}(n_1,n_2))$.

This expression is to be compared with the projection onto ${\cal H}_\gamma$
of the operator of \cite{1.4}
\ba \label{a.6}
\hat{A}'_{S,\gamma}&=&\frac{1}{2}\sum_{n_2}\{
\sum_{n_1\not=-N+kM}\sqrt{[J^{3,up}_j-J^{3,down}_j]^2(X(\k{v}(n_1,n_2))}
\nonumber\\
&& +\sum_{n_1=-N+kM}
\sqrt{[J^{3,up}_j-J^{3,down}_j
+J^{1,left}_j-J^{3,right}_j]^2(X(\k{v}(n_1,n_2))}
\}
\ea
where $J^e_j=i\ell_p^2/2\mbox{tr}((\tau_j h_e)^T\partial/(\partial 
h_e))$ and $J^{I,*}(v)_j=J^{e_I(v)^*}_j$ where $e_I(v)^*$ is the segment 
of $e_I(v)$ outgoing from the intersection of $e_I(v)$ with $S$ into the 
direction $*=up,down,left,right$ and $e_I(v)=X(\k{e}_I(\k{v}))$. A simple
calculation reveals that for 
$e=(e^{down})^{-1}\circ e^{up}$ or $e=(e^{left})^{-1}\circ e^{right}$ we 
have $J^{I,*}_j(v)f(h_e)=a^2 \hat{P}^I(v)_j f(h_e)$ if $*=up,right$ and
$J^{I,*}(v)f(h_e)=-a^2 O_{jk}(h_e) \hat{P}^I(v)_k f(h_e)$ if $*=down,left$ 
where the orthogonal matrix $O_{jk}(h)$ is defined by $\mbox{Ad}_h(\tau_j)
=O_{jk}(h)\tau_k$. It follows that
\ba \label{a.7}
&& \hat{A}'_{S,\gamma} = \frac{a^2}{2}\sum_{n_2}\{
\sum_{n_1\not=-N+kM}
\sqrt{[(\delta_{jk}+O_{jk}(h_{e^3(v(n_1,n_2))}))
\hat{P}^3_k(v(n_1,n_2)]^2}
\\
&+& \sum_{n_1=-N+kM}
\sqrt{[(\delta_{jk}+O_{jk}(h_{e^3(v(n_1,n_2))}))
\hat{P}^3_k(v(n_1,n_2)-(\delta_{jk}+O_{jk}(h_{e^1(v(n_1,n_2))}))
\hat{P}^1_k(v(n_1,n_2)]^2}
\}
\nonumber
\ea

Let us test these expressions in a coherent state for flat data 
$m=(A_a^j=0,E^a_j=\delta^a_j)$. Notice that then the parameter
$\epsilon$ is really the edge length determined by $m$. Then 
$h_{e^I(v)}=1,P^I_j(v)=\delta^I_j\epsilon^2$ and the Ehrenfest theorems
proved in \cite{1.3} reveal that the zeroth order in $\hbar$ of the 
expectation values is given by 
\ba \label{a.8}
&& \lim_{t\to 0} 
<\psi_{\gamma,m},\hat{A}_{S,\gamma}
\psi_{\gamma,m}>
=(2N\epsilon)^2\sqrt{1+\tau^2}
\nonumber\\
&& \lim_{t\to 0} 
<\psi_{\gamma,m},\hat{A}'_{S,\gamma}
\psi_{\gamma,m}>
=(2N\epsilon)^2[1+\tau(\sqrt{2}-1)]
\ea
The value in the first line of (\ref{a.8}) 
is in fact the exact classical area of the surface while the second line 
is off (equal or bigger than) that value. {\it This is the staircase 
problem}. We get coincident
results only for $\tau=0,1$ corresponding to $M=\infty,0$. The relative 
mistake is zero for $M=1$, then reaches its maximum of 8\% at $M=2$
and then monotonously decreases as $1/M$ to zero. An error of
eight percent is inacceptible which is why we are forced to use  
$\hat{A}_S$ instead of $\hat{A}'_S$ which is guaranteed to give results of 
sufficiently high accuracy. Notice that the dual faces of the graph under 
consideration in this example {\it are not at all tangential to the surface
$S$} and still we get a good (even exact) result. 

Yet, this example was special in the sense that the coefficients
$\mu_k,n_I(k)$ were independent of $k$ so that one would have gotten the 
same result even if there was only one cube needed (replace $N$ by $1$
and $\epsilon$ by $N\epsilon$) which is due to the fact that in this 
case (\ref{a.2}) gives the exact classical value.
If the surface would not be a plane, then the coefficients would vary 
and (\ref{a.2}) no longer gives the exact classical value. If the 
surface $S$ wiggles at a scale much lower than $\epsilon$ then 
(\ref{a.2}) will be a 
very bad approximation. In that case we can improve the result by increasing
the number of edges of the lattice and thus this error is a {\it finite size
effect}. Next, in replacing (\ref{a.1}) by (\ref{a.2}) it was crucial 
that $m$ does not vary too much on an embedded edge of the graph or the dual
embedded face. An error due to this is related to the curvature of $m$ 
and will be called {\it a curvature effect}.

In what follows we will exclude finite size effects by the assumption that
the surfaces that we are interested in are wiggling at most at the scale of 
the graph. More precisely, the $S_k=X(\k{S}_k)$ are allowed not to lie at all 
inside the dual faces of the embedded graph but the tangent space should
be approximately constant over $S_k$. For instance, we allow for a surface
which is parallel to the $t^2$ direction as above but such that $\tau$
jumps between $\pm 1$ from cube to cube (its projection into the $t^1,t^3$
plane is a zig-zag curve with triangle edge length $\sqrt{2}\epsilon$).
However, we do not allow $\tau$ to jump within one cube.
We are then left with curvature effects to which we turn in the next
section.

\section{Minimization of Fluctuations through Optimization of Scales} 
\label{sb}

In this section we will sketch the computation of fluctuations in our scheme
leading to a fixing of the free parameters $a,\epsilon$ through 
optimization.\\

Given the fact that we are in a gauge field theory context, it is 
motivated to consider as an elementary set of classical continuum
observables the non-Abelean analogues of the electric and magnetic fluxes
through embedded surfaces $S$. At this point we could say that we measure 
only the quantum fluxes associated with the dual faces of a given graph 
since this information is enough in order to reconstruct any other 
classical observable in the fashion outlined in the previous section 
\ref{sa}. In that case and after normal ordering there would be no
classical or normal ordering error at all. However, one would like to be 
more general and consider fluxes through arbitrary surfaces and 
not only elementary ones because one would like to measure these more general
obervables directly at the quantum level and not only through an indirect
procedure using a classical formula. Actually, if the Gauss law 
$\partial_a E^a_j=0$ holds and due to the Bianchi identity, we can replace 
$S$ by any surface 
$S'$ built entirely from dual faces of the polyhedronal decomposition 
such that $S-S'=\partial R$ is the boundary of a region $R$
(possibly up to some neglible piece at the boundary of $S$) without changing 
the classical 
flux functions. Thus, for the fluxes (on the constraint surface of the Gauss
constraint) we could actually keep the discussion at the level of 
surfaces built from elementary faces which would simplify the discussion,
however, we will not do that for the sake of generality and in view of 
applications to other sets of basic variables which are not fluxes and 
for which an off-shell approximation might be desirable.

In order to keep the presentation at an elementary level we will replace
$SU(2)$ by $U(1)^3$ as otherwise the non-Abeleaness blows up the effort 
to obtain all the 
estimates by an order of magnitude (see especially the fourth reference 
in \cite{1.3}) without changing the end result. We are thus concerned with 
the following classical (kinematical) observables
\be \label{b.1}
B_j(S)=\int_S F_j(x) \mbox{ and } E_j(S)=\int_S (\ast E_j)(x) 
\ee
where $F_j$ is the curvature two-form of $A^j$. Notice that $F_j,E_j$ 
are gauge invariant. In the non-Abelean case one would need to choose 
a point $p$ inside $S$ and a system of paths between $p$ and any other
point $x\in S$ and integrate instead of $\ast E_j(x), F_j(x)$ their
image under the adjoint action of $SU(2)$ evaluated at the holonomies
along those paths in order to obtain a gauge-covariant result. The 
non-Abelean case will be discussed in more detail elsewhere.

Using the same notation as in the previous section we have a model
graph $\k{\gamma}$ of cubic topology and a model surface $\k{S}$
of square topology which are embedded into $\Sigma$ via embeddings
$X$ and $Y=X\circ \vec{t}$ respectively so that 
\be \label{b.2}
B_j(S)=\int_{\k{S}} du dv 
[(n_a^I(\vec{t})B^a_j(X(\vec{t}))_{\vec{t}=\vec{t}(u,v)}n_I(u,v)
\ee
and similar for $E_j(S)$ where $B^a_j=\frac{1}{2}\epsilon^{abc}F_{bc}^j$.
Using the same dual polyhedronal decomposition and choices of paths as in 
the previous section we arrive at the classical substitute functions
\ba \label{b.3}
B_{j,\gamma}(S) & = & \sum_k \mu_k n_I(u_k,v_k) 
\frac{h^j_{\alpha^I(v_k)}-(h^j_{\alpha^I(v_k)})^{-1}}{2i}
\nonumber\\
E_{j,\gamma}(S) & = a^2 & \sum_k \mu_k n_I(u_k,v_k) 
P^I_j(v_k)
\ea
where $\alpha^I(v)=X(\k{\alpha}_I(\k{v}))$ is the image under $X$ 
of the model plaquette loop 
$\k{\alpha}_I(\k(v))=
\k{e}_J(\k{v})\circ \k{e}_K(\k{v}+b_J)^{-1}\circ
\k{e}_J(\k{v}+b_K)^{-1}\circ
\k{e}_K(\k{v})^{-1}$ on 
$\k{\gamma}$ in the $J,K$ plane with $\epsilon_{IJK}=1$ and with vertex
at $\k{v}=X^{-1}(v)$. Here, $b_I$ denotes the standard orthonormal
basis in $\Rl^3$. Notice that the path system $\Pi_\gamma$ is actually not 
needed
in order to define $P^I_j(v)$ since $U(1)^3$ is Abelean. Due to the relation 
\be \label{b.4}
g^j_I(v)=e^{P^I_j(v)}h^j_{e_I(v)}
\ee
it is possible to replace (\ref{b.3}) by functions which are linear
combinations of products of the 
$g^j_I(v),(g^j_I(v))^{-1},\overline{g^j_I(v)},\overline{(g^j_I(v))^{-1}}$.
These functions are the classical analogues of the creation operators
$\hat{g}^j_I(v),(\hat{g}^j_I(v))^{-1}$ and annihilation
operators $(\hat{g}^j_I(v))^\dagger,((\hat{g}^j_I(v))^{-1})^\dagger$
of section \ref{s3} and with the help of these functions 
it becomes possible to normal order the corresponding operators in the 
quantum theory. For example, $g^j_I(v)+\overline{(g^j_I(v))^{-1})}=
h^j_I(v)(1+O((\frac{\epsilon}{a})^4))$ and 
$\overline{g^j_I(v)}g^j_I(v)-\overline{(g^j_I(v))^{-1}}(g^j_I(v))^{-1}
=4P^I_j(v)(1+O((\frac{\epsilon}{a})^4))$. By using (finitely many) 
higher powers of
these four functions one can suppress the subleading terms to any
desired order in $(\epsilon/a)$.

We can then quantize (\ref{b.3}) as
\ba \label{b.5}
\hat{B}_{j,\gamma}(S) & := & \sum_k \mu_k n_I(u_k,v_k) \;
:\frac{\hat{h}^j_{\alpha^I(v_k)}
-(\hat{h}^j_{\alpha^I(v_k)})^{-1}}{2i}:
\nonumber\\
\hat{E}_{j,\gamma}(S) & = a^2 & \sum_k \mu_k n_I(u_k,v_k) \;
:\hat{P}^I_j(v_k):
\ea
where the normal ordering symbol is to be understood after substituting
the just mentioned approximations up to $(\epsilon/a)^{2n}$ for some
desired $n$. There is then no normal ordering error. 

Next, we are  
concerned with the classical error given by
\ba \label{b.6}
E_j(S) (\Delta E_j(S))_{class}&\le& 
a^2 \sum_k \mu_k |n_I(u_k,v_k)[:P^I_j(v_k):-P^I_j(v_k)]|
\nonumber\\
&& +|E_j(S)-\sum_k \mu_k n_I(u_k,v_k) E_j(S^I(v_k)|
\ea
where $S^I(v)$ is the image under $X$ of the surface $\k{S}^I(\k{v})$ dual
to $\k{e}_I(\k{v})$ and similar for $(\Delta B_j(S))_{class}$. The normal
ordering symbols in (\ref{b.6}) just mena that one shoul replace 
the functions inside the normal ordering symbols by the above mentioned
functions of the $g^j_I(v)$.
Let us estimate these terms. If we denote by $A_S(E)$ the classical area
of $S$ determined by $E$ then the number of terms in the sum $\sum_k$
will be of the order of $A_S(E)/\epsilon^2$ and so the first term can be 
bounded by $A_S(E)(\tilde{E}^j_S)^n (\epsilon/a)^{2(n-1)}$ where 
$\tilde{E}^j_S$ is the maximum of $E_j(S')/A_E(S')$ as we vary $S'\subset S$.  
For the second term we employ the Euler-MacLaurin estimates \cite{b.1}
for the numerical difference between an integral and a Riemann sum.
We will use only the coarsest estimate resulting from the identity 
\be \label{b.7}
\int_y^{y+N\epsilon} dx F(x)-\epsilon[\frac{F(y)+F(y+N\epsilon)}{2}
+\sum_{k=1}^{N-1} F(y+k\epsilon)]
=\frac{\epsilon^3}{2}\int_0^1 dt \phi_2(t)[\sum_{k=0}^{N-1} 
F^{\prime\prime}(y+(m+t)\epsilon)]
\ee
where $\phi_2(t)=t^2-t$ is the Bernoulli polynomial of the second degree
and $F$ is a twice differentiable function on the interval 
$[y,y+N\epsilon)]$. Using $\phi_2(t)\le 1/4$ for $0\le t\le 1$ the right
hand side of (\ref{b.7}) can be bounded from above by
$\epsilon^2/8\int_y^{y+N\epsilon} dx |F^{\prime\prime}(x)|$ and thus involves
the {\it second derivative} of $F$. Let us introduce the {\it curvature
scale} 
\be \label{b.8}
\frac{1}{L_F^2}:=
|\frac{\frac{\epsilon}{2}\int_0^1 dt \phi_2(t)[\sum_{k=0}^{N-1} 
F^{\prime\prime}(y+(m+t)\epsilon)]}{\int_y^{y+N\epsilon} dx F(x)}|
\ee
Using this formula and iterating it since we are dealing with a 
two-dimensional integral we find that the second term can be bounded
by $|E_j(S)|\epsilon^4/L_E^4$ where $L_E$ is a lower bound on the 
{\it electric curvature radius} of the phase space point $m$ and thus
independent of $j,S$. 

We will now make the assumption that $\epsilon\ll a,L_E$ and that $n$
is sufficiently high so that the first term in (\ref{b.6}) is sub-leading
(since we will relate $a$ to $L_E$ later on, this is a self-consistency
assumption). Then we obtain (estimates are in orders, so we neglect
numerical factors of order one)
\be \label{b.9}
(\Delta E_j(S))_{class}\le \frac{\epsilon^4}{L_E^4} \mbox{ and }
(\Delta B_j(S))_{class}\le \frac{\epsilon^4}{L_B^4}
\ee
where the magnetic curvature radius will be of the same order as $L_E$
since both types of curvature come from the same four-dimensional curvature
tensor.

Finally we are left with the quantum mechanical error. Notice that 
the operators (\ref{b.5}) are of the form 
$\hat{O}=\sum_k \hat{O}_k$ where the $\hat{O}_k$ mutually commute for
the electric flux while for the magnetic flux the next neighbour terms
are non-commuting. However, the number of these non-commuting terms 
per vertex is of the same order as the number of the $\hat{O}_k$ and 
since we are just concerned with orders of 
magnitude here their contribution to the following estimates gives just a 
multiplicative numerical factor of order one which we do not display.
We can thus assume that all the $\hat{O}_k$ are mutually commuting and 
obtain for the fluctuations in the coherent state
$\psi_{\k{\gamma},\Sigma,X,m}$
\be \label{b.10}
<\hat{O}>^2(\Delta O)(m)_{quantum}^2=\sum_k [<\hat{O}_k^2>-<\hat{O}_k>^2]
\le t \frac{A_E(S)}{\epsilon^2}
\ee
where the Ehrenfest theorems for $\hat{O}_k=\hat{P}^I_j(v),\hat{h}^j_I(v)$
of the third reference in \cite{1.3} have been 
used in the estimate. For the total error we therefore find
\ba \label{b.11}
(\Delta E_j(S))_{total}(m)^2& \le& t\frac{a^4 A_E(S)}{\epsilon^2 E_j(S)^2} 
+\frac{\epsilon^8}{L_E^8}
\approx \frac{\ell_p^2 a^2}{\epsilon^2 A_S(E) \bar{E}_S}
+\frac{\epsilon^8}{L_E^8}
\nonumber\\
(\Delta B_j(S))_{total}(m)^2& \le& t\frac{A_E(S)}{\epsilon^2 B_j(S)^2} 
+\frac{\epsilon^8}{L_B^8}\approx \frac{\ell_p^2 L_B^4}{a^2 \epsilon^2 A_S(E)}
+\frac{\epsilon^8}{L_B^8}
\ea
where $\bar{E}_S=\sqrt{E_j(S)^2}/A_E(S)$ is of order unity and in the last 
line we have 
used the approximate identity $B_j(S)\approx A_E(S)/L_B^2$. 

In order to fix the parameters $\epsilon,a$ we notice that by assumption
$A_S(E)\ge \epsilon^2$ ($N=A_S(E)/\epsilon^2$ is the number of contributing
faces and gives the law of large numbers $\propto 1/\sqrt{N}$ for the 
quantum mechanical fluctutations). Thus we can further estimate
\ba \label{b.12}
(\Delta E_j(S))(m)^2& \le& 
\frac{\ell_p^2 a^2}{\epsilon^4 \bar{E}_S}
+\frac{\epsilon^8}{L_E^8}
\nonumber\\
(\Delta B_j(S))(m)^2& \le& 
\frac{\ell_p^2 L_B^4}{a^2 \epsilon^4}
+\frac{\epsilon^8}{L_B^8}
\ea
and we require that these two errors be equal (yielding an ``unquenchedïï
uncertainty relation for the fluxes) and minimal. One can solve 
$\Delta E=\Delta B$ analytically for $a^2$ resulting in 
$a^2=\alpha x^3+\sqrt{(\alpha x^3)^2+L_B^4 \bar{E}_S}$ where 
$\alpha=\bar{E}_S(1/L_B^8-1/L_E^8)/(2\ell_p^2)$ and $x=\epsilon^4$. The 
remaining optimization
problem can then only be solved numerically. We will not treat the most
general case in this paper but make the physical assumption that 
$L_E\approx L_M\approx L$ so that $a^2\approx\sqrt{\bar{E}_S} L^2$. 
Notice that provided $\epsilon\ll L$ our self-consistency 
assumption from above is indeed satisfied. The optimization then
yields an absolute minumum at 
$\epsilon^{12}=\ell_p^2 L^{10}/(2\sqrt{\bar{E}_S})$. Thus, $\epsilon$
is locked at a {\it geometric mean} of the microscopic and macroscopic 
scale. We verify that indeed  
$\ell_p\ll \epsilon\ll L$ as long as $\ell_p\ll L$,
finishing up our self-consistency check. The total 
fluctuation becomes up to a numerical factor equal to 
$(\frac{\ell_p}{L \bar{E}_S})^{4/3}\le 
(\frac{\ell_p}{L_0 \bar{E}_S})^{4/3}$ where $L_0$ is the phase space 
bound mentioned in the main text and depends on $\cal M$ but not on $m$.
Locking $a$ at $L_0$ from the outset would lead to qualitatively similar
results.\\
\\
We finish this section with a couple of remarks :\\
1.) Our analysis has optimized only an upper bound, it may well be that
one can get even better estimates depending on $m$.\\
2.) One may wonder why the total fluctuation of our operator should depend 
on a classical error which then leads to an optimization problem in 
$\epsilon$ which one is not used to from electrodynamics. However, there
actually the same effect happens \cite{1.12} : If one considers in free
Maxwell theory the fluctuations of electric and magnetic flux
operators then one simply finds infinity. This is due to the fact that
in the corresponding Fock represenation one has to smear the fields over
three dimensional regions rather than two-dimensional surfaces in order to 
arrive at a well-defined operator. Thus, in order to make the fluctuation
calculation well-defined one has to fatten the surfaces by a transversal 
thickness $\epsilon$ (which is a regularization comparable to our replacing
continuum functions by discretized objects) and then divides the fattened 
flux by $\epsilon$. When one now computes the 
fluctuation of these operators compared to exact clssical flux value 
one finds a similar competition between the quantum mechanical fluctuation
blowing up as $\epsilon\to 0$ and the classical error vanishing as 
$\epsilon\to 0$. The optimization thus takes place since one is interested
in measuring classcical obervables quantum mechanically which are smeared
by singular smearing functions. Whether this is a sensible thing to do or
not is a {\it physical} question. If one insists to measure fluxes quantum
mechanically then one has no choice and fluxes are precisely the kind of
objects that are natural in our non-Fock representation.

\section{The Infinite Spin Chain}
\label{sc}

In this final appendix we consider a trivial but instructive physical
system in order to exemplify infinite tensor product concepts, namely
an infinite chain of uncoupled spin degrees of freedom.\\
\\
Our label set will be the integers $I=\Zl$ and for each $n\in \Zl$ 
we have the Hilbert space ${\cal H}_n=\Cl^2$ with standard
inner product $<f_n,f_n'>_n=\bar{f}_n^+ f_n^{+\prime}
+\bar{f}_n^- f_n^{-\prime}$. In each Hilbert space 
we have the standard orthogonal basis of vectors $e_n^\pm$ and spin 
operators $\sigma_n=\sigma_3$ (Pauli matrix) so that $\sigma_n e^\pm_n 
=\pm e^\pm_n$ corresponds to spin up/down. We also have ladder operators
$\sigma^\pm_n=\frac{1}{2}[\sigma_1\pm i\sigma_2]$ so that 
$\sigma_n^\pm e_n^\pm=0,\; \sigma_n^\pm e_n^\mp=e^\pm_n$. Consider
the positive semi-definite, self-adjoint Hamiltonian 
\be \label{c.1}
\hat{H}:=\frac{1}{2}\sum_n [1+\sigma_n]=
\sum_n (\sigma^-_n)^\dagger \sigma^-_n
\ee
on the ITP Hilbert space ${\cal H}^\otimes =\otimes_n {\cal H}_n$ which
is non-separabe even though each ${\cal H}_n$ has finite dimension two.

We will first consider a $C_0$ vector $\otimes_f$ with $||f_n||_n=1$ and a 
second one $\otimes_{f'}$ with $f'_n=-f_n$. Are the correspeonding 
$C_0$ sequences in the same strong
(weak) eqivalence class ? Since $<f_n,f'_n>_n=-1$ we see that
$\sum_n |<f_n,f'_n>_n-1|=\sum_n 2=\infty$ but
$\sum_n ||<f_n,f'_n>_n|-1|=\sum_n 0=0$, thus they are in different 
strong classes within the same weak class. In fact, the unitary 
operator $\hat{U}$ on ${\cal H}^\otimes$ defined densely on arbitrary 
$C_0$ vectors by $\hat{U}\otimes_g=\otimes_{g'}$ with $g_n'=-g_n$
maps the two unit $C_0$ vectors into each other and thus the strong 
equivalence class Hilbert spaces built from them will be unitarily
equivalent subspaces of the whole ITP. Notice that indeed
$<\otimes_f,\otimes_{f'}>=\prod_n (-1)^n=0$ since the product of numbers
$z_n=-1$ is only quasi-convergent.

Consider now the unit $C_0$ vectors $\Omega^\pm:=\otimes_{f^\pm}$ with 
$f^\pm_n:=e^\pm_n$ (the total spin up/down state). We have 
$\sum_n ||<f^+_n,f^-_n>_n|-1|=\sum_n 1=\infty$, thus they are in different 
weak classes. It is true that the {\it non-local operator}
$\hat{A}:=\otimes_n \sigma_n^+$ maps $\Omega^-$ into $\Omega^+$, however,
$\hat{A}$ is not a unitary operator as one easily verifies by computing
the norm of $\hat{A}\otimes_f$ with $f^+_n\not=0$ for at least one $n$.
In fact, $\Omega^-$ is the {\it ground state} of $\hat{H}$ while  
$\Omega^+$ is an infinite energy (infinitely excited) state for $\hat{H}$
and thus their strong equivalence class Hilbert spaces describe drastically
different physics. It is therfore not to be expected on physical grounds 
that these representations should be unitarily equivalent.

Finally, it is easy to see that the orthonormal ``spin-network" $C_0$ vector 
sytem $\otimes_{\{\alpha_n\}}:=\otimes_n e_n^{\alpha_n}$ with $\alpha_n=\pm$
is not a basis : For instance the $C_0$ vector with tensor product factors
$f_n:=\frac{1}{\sqrt{2}}[e_n^+ + e_n^-]$ is orthonormal to all of them,
a nice example in which one sees that the unrestricted associative law
is false.

With the help of simple systems like the infinite spin chain one can study a 
lot of 
more phenomena of the ITP such as the occurrence of von Neumann algebra
factor types $I_\infty,II_1,III_s,\;s\in(0,1)$. The von Neumann algebra
of local operators ${\cal R}^\otimes$ considered in the main text is of 
that type if we restrict it to the strong equivalence class Hilbert spaces 
defined by the cyclic vectors 
\be \label{c.2}
\Omega_s:=\otimes_n [\sqrt{\frac{1+s}{2}} \sigma^+_n
+\sqrt{\frac{1-s}{2}} \sigma^-_n]
\ee
of the ITP Hilbert space ${\cal H}^\otimes:=\otimes_n [\Cl^2\otimes\Cl^2]$ 
for $s\in\{1\},\{0\},(0,1)$ respectively. These representations can be 
interpreted as zero, infinite and finite temperature representations 
respectively, see \cite{4.4} for details.

\end{appendix}


\begin{thebibliography}{99}


\parskip -5pt

\bibitem{1.2} 
A. Ashtekar, ``Non-Perturbative Canonical Gravity''
Lectures notes prepared in collaboration with R.S. Tate, World Scientific,
Singapore, 1991; in {\it Gravitation and Quantization}, B. Julia (ed)
Elsevier, Amsterdam, 1995\\
C. Rovelli, ``Loop Quantum Gravity", Review written for
electronic journal ``Living Reviews", gr-qc/9710008;
``Strings, Loops and Others : a critical Survey of the
present Approaches to Quantum Gravity", plenary lecture given at
15th Intl. Conf. on Gen. Rel. and Gravitation (GR15), Pune, India,
Dec 16-21, 1997, gr-qc/9803024\\
A. Ashtekar, C. Beetle, O. Dreyer, S. Fairhurst, B. Krishnan, 
J. Lewandowski, J. Wisniewski, Phys. Rev. Lett. {\bf 85} 
(2000) 3564-3567, gr-qc/0006006    

\bibitem{1.1} 
R. Gambini, J. Pullin, Phys. Rev. {\bf D59} (1999) 124021\\
J. Alfaro, H. A. Morales-Tecotl, L. F. Urrutia, 
Phys. Rev. Lett. {\bf 84} (2000) 2318-2321\\
M. Varadarajan, J.-A. Zapata, 
Class. Quantum Grav. {\bf 17} (2000) 4085, gr-qc/0001040\\
M. Arnsdorf, ``Approximating Connections in Loop Quantum Gravity'',
gr-qc/9910084\\
M. Arnsdorf, S. Gupta, Nucl. Phys. {\bf B577} (2000) 529\\
A. Corichi, ``A Gaussian Weave for Kinematical Loop Quantum Gravity'',
gr-qc/0006067

\bibitem{1.3} 
T. Thiemann, ``Symplectic Structures and Continuum Lattice 
Formulations of Gauge Field Theories'', hep-th/0005232;
``Gauge Field Theory Coherent States (GCS) : I.
General Framework'', hep-th/0005233\\
T. Thiemann, O. Winkler, ``Gauge Field Theory Coherent States (GCS) : II.
Peakedness Properties'', hep-th/0005237;
``Gauge Field Theory Coherent States (GCS) : III.
Ehrenfest Theorems'', hep-th/0005234;
``Gauge Field Theory Coherent States (GCS) : IV.
Infinite Tensor Product and Thermodynamic Limit'', hep-th/0005235

\bibitem{1.4} C. Rovelli, L. Smolin, 
Nucl. Phys. B {\bf 442} (1995) 593; Erratum : Nucl. Phys. B {\bf 456} 
(1995) 734 \\
A. Ashtekar, J. Lewandowski, 
Class. Quantum Grav. {\bf 14} A55-81 (1997)


\bibitem{1.5} A. Ashtekar, L. Bombelli, 
"Statistical Geometry and Semiclassical Quantum Gravity", to appear

\bibitem{1.6} C. Itzykson, J.-M. Drouffe, ``Statistical Field Theory'',
vol. 2, Cambridge University Press, Cambridge, 1997

\bibitem{1.5a} A. Ashtekar, C. Rovelli, L. Smolin, Phys. 
Rev. Lett. {\bf 69} (1992) 237

\bibitem{1.7} C. Rovelli, L. Smolin,
Phys. Rev. {\bf D52} (1995) 5743\\
J. Baez,
Adv. Math. {\bf 117} (1996) 253

\bibitem{1.8} T. Thiemann, ``Quantum Spin Dynamics (QSD) VIII. : The 
Classical Limit'', in preparation

\bibitem{1.9} 
A. Ashtekar, J. Lewandowski, D. Marolf, J.
Mour\~ao, T. Thiemann,
Journ. Math. Phys. {\bf 36} (1995) 6456-6493, [gr-qc/9504018]

\bibitem{1.10} 
T. Thiemann,
Physics Letters B {\bf 380} (1996) 257-264, [gr-qc/960688];
Class. Quantum Grav. {\bf 15} (1998) 839-73, [gr-qc/9606089];
Class. Quantum Grav. {\bf 15} (1998) 875-905, [gr-qc/9606090];
Class. Quantum Gravity {\bf 15} (1998) 1207-47, [gr-qc/9705017];
Class. Quantum Grav. {\bf 15} (1998) 1249-80, [gr-qc/9705018];
Class. Quantum Grav. {\bf 15} (1998) 1281-1314, [gr-qc/9705019];
Class. Quantum Grav. {\bf 15} (1998) 1487-1512, [gr-qc/9705021]

\bibitem{1.11} 
A. Ashtekar, J. Lewandowski, D. Marolf, J. Mour\~ao, T. Thiemann,
Journ. Funct. Analysis {\bf 135} (1996) 519-551, [gr-qc/9412014]

\bibitem{1.12} A. Ashtekar, L. Bombelli, O. Dreyer, S. Fairhurst, A. Gupta,
H. Sahlmann, T. Thiemann, O. Winkler, ``Kinematical versus
Dynamical Coherent States'', in preparation

\bibitem{1.13} 
H. Sahlmann, T. Thiemann, O. Winkler, ``Canonical Quantum General Relativity
and Algebraic Graph Theory'', in preparation

\bibitem{1.14} N. Biggs, ``Algebraic Graph Theory'', Cambridge University
Press, 2nd ed., Cambridge, 1993\\
D. Stauffer, A. Aharony, ``Introduction to Percolation Theory'',
Taylor and Francis, 2nd ed., London, 1994\\
D. M. Cvetovic, M. Doob, H. Sachs, ``Spectra of Graphs'', Academic Press,
New York, 1979 

\bibitem{3.1} B. C. Hall, Journ. Funct. Analysis {\bf 122} (1994) 103;
Comm. Math. Phys. {\bf 184} (1997) 233-250

\bibitem{4.1} A. Ashtekar, C. J. Isham,
Class. Quantum Grav. {\bf 9} (1992) 1433\\
A. Ashtekar, J. Lewandowski, ``Representation
theory of analytic holonomy $C^\star$ algebras'', in ``Knots and
quantum gravity, J. Baez (ed), Oxford University Press, Oxford 1994;
Journ. Geo. Physics {\bf 17} (1995) 191;
J. Math. Phys. {\bf 36}, (1995) 2170

\bibitem{4.2} J. Baez, S. Sawin, ``Functional Integration on Spaces of
Connections'', q-alg/9507023;
``Diffeomorphism Invariant Spin-Network States'', q-alg/9708005\\
J. Lewandowski, T. Thiemann,
Class. Quantum Grav. {\bf 16} (1999) 2299-2322, gr-qc/9901015

\bibitem{4.3} J. von Neumann, Comp. Math. {\bf 6} (1938) 1-77

\bibitem{4.4} W. Thirring, ``Lehrbuch der Mathematischen Physik'', vol. 4, 
Springer-Verlag, Wien, 1994

\bibitem{4.5} A. Connes, ``Noncommutative Geometry'', Academic Press, 1994

\bibitem{4.6} R. Haag, ``Local Quantum Physics'', 2nd ed., Springer Verlag,
Berlin, 1996

\bibitem{4.7} Y. Yamasaki, ``Measures on Infinite Dimensional Spaces'',  
World Scientific, Singapore, 1985

\bibitem{4.8} 
H. Sahlmann, T. Thiemann, O. Winkler ``Gauge Field Theories Coherent States
(GCS) : VI. Photons and Gravitatons Propagating on Quantum Spacetimes'',
in preparation

\bibitem{5.1} M. Bojowald, H. Kastrup, Class. Quantum Grav. {\bf 17} (2000)
3009 [hep-th/9907042];
Class. Quantum Grav. {\bf 17} (2000) 3044 [hep-th/9907043]\\
M. Bojowald, J. Math. Phys. {\bf 41} (2000) 4313 [hep-th/9908170];
Class. Quantum Grav. {\bf 17} (2000) 1489 [gr-qc/9910103];
Class. Quantum Grav. {\bf 17} (2000) 1509 [gr-qc/9910104];
[gr-qc/0008052];
[gr-qc/0008053];

\bibitem{5.2} A. Ashtekar, J. Lewandowski, 
Adv. Theo. Math. Phys. {\bf 1} (1997) 388





\bibitem{b.1} E.T. Whittaker, G. N. Watson, ``A Course of Modern Analysis'',
chapter VII, Cambridge University Press, Cambridge, 1969\\
H. Jefferys, B. S. Jeffreys, ``Methods of Mathematical Physics'',
Cambridge University Press, Cambridge, 1992


\end{thebibliography}
\end{document}